\documentclass[11pt]{article}

\usepackage[final]{acl}

\usepackage{times}
\usepackage{latexsym}
\usepackage{booktabs}
\usepackage[T1]{fontenc}
\usepackage[utf8]{inputenc}
\usepackage{microtype}
\usepackage{inconsolata}
\usepackage{graphicx}
\usepackage{amsmath}
\usepackage{amssymb}
\usepackage{xcolor}    % 颜色
\usepackage{multirow} 
\usepackage{colortbl}  % 表格颜色
\usepackage{tabularx}
\usepackage{tikz} 
\usepackage[linesnumbered,ruled,vlined]{algorithm2e}
\usepackage{listings}
\usepackage{tcolorbox}
\tcbuselibrary{listings,skins,breakable}
\usepackage{caption}
\usepackage{enumitem}
\usepackage{arydshln}
\usepackage{colortbl}
\usepackage{hyperref}
\usepackage{pifont}

\newcommand{\cmark}{\ding{51}}%
\newcommand{\xmark}{\ding{55}}%
\definecolor{bg_fail}{RGB}{255, 240, 240} % Light Red for low confidence
\definecolor{bg_success}{RGB}{240, 255, 240} % Light Green for high confidence
\definecolor{bg_process}{RGB}{248, 248, 255} % Light Blue for neutral process
\definecolor{rowgray}{rgb}{0.95, 0.95, 0.95}
\definecolor{cfactual}{HTML}{9CC3D5}      % Misty blue
\definecolor{cinferential}{HTML}{C0AED9}  % Soft lilac
\definecolor{ctemporal}{HTML}{F0D1A2}     % Warm pastel ochre
\definecolor{csummarization}{HTML}{C6D7B9}% Muted sage
\definecolor{cacoustic}{HTML}{F3C5C5}     % Blush rose
\definecolor{c_gt}{HTML}{4E79A7}     % Blush rose

\DeclareRobustCommand{\categorylabel}[2]{%
    \begin{tikzpicture}[baseline=(X.base)]
        \node[
            fill=#1,                % 背景填充色
            rounded corners=3pt,    % 圆角大小 (数值越大越圆)
            inner sep=3pt,          % 文字与边框的距离 (Padding)
            text=black,             % 文字颜色
            font=\bfseries\small,   % 字体加粗、变小
            anchor=center             % 对齐方式
        ] (X) {#2};
    \end{tikzpicture}%
}

\newtcolorbox{qabox}[1]{
  enhanced,
  breakable,
  colback=white,
  colframe=black!60,
  boxrule=0.6pt,
  arc=2pt,
  left=6pt,right=6pt,top=5pt,bottom=5pt,
  width=\linewidth,
  title=\textbf{#1},
  fonttitle=\normalsize,
}
\newtcolorbox{promptbox}[1][]{
  colback=gray!5!white,    % 浅灰色背景
  colframe=gray!75!black,  % 深灰色边框
  title=#1,                % 标题
  fonttitle=\bfseries,     % 标题加粗
  fontupper=\small\ttfamily, % 内容使用打字机字体(代码风)且稍小
  boxrule=0.8pt,
  arc=2pt,
  left=5pt, right=5pt, top=5pt, bottom=5pt,
  sharp corners=south,     % 可选：底部直角
}
\newcommand{\qafield}[2]{\noindent\textbf{#1:}~#2\par}

\title{Don't Just Listen, Try Planning: Graph-based Retrieval-Generation Agent for Long-form Audio Meeting Understanding}

\author{
 \textbf{Quanwei Tang\textsuperscript{1}},
 \textbf{Dong Zhang\textsuperscript{1,2*\footnote{* Corresponding Author}}},
 \textbf{Shoushan Li\textsuperscript{1}}, and
 \textbf{Guodong Zhou\textsuperscript{1}}
\\
\\
 \textsuperscript{1}School of Computer Science \& Technology, NLP Lab, Soochow University, China \\  \textsuperscript{2}Jiangsu Key Lab of Language Computing, Suzhou. \\
 \small{
   \href{mailto: dzhang@suda.edu.cn}{ dzhang@suda.edu.cn}
 }
}

\begin{document}
\maketitle
\begin{abstract}
While long-form audio meeting understanding (LAMU) is garnering growing attention, task-specific question answering (QA) datasets remain scarce. Existing speech QA paradigms and state-of-the-art Speech LLMs suffer from acoustic information loss and poor long-term context memory. To address these issues, we construct the LongAudioQA dataset and propose the GRGA model,  which models heterogeneous audio features into a multi-dimensional graph and leverages agent planning for retrieval and answer generation. ~\href{https://github.com/tangquanwei/GRGA}{GitHub} for data and code.
% Long-form audio meeting understanding (LAMU) is gaining attention, but dedicated question answering (QA) datasets are lacking. Previous tailored speech QA and existing Speech LLMs suffer from acoustic information loss and poor long-term context memory. We construct the LongAudioQA dataset and propose the GRGA model, which models heterogeneous audio features into a multi-dimensional graph and leverages agent planning for retrieval and answer generation, effectively addressing existing limitations.~\href{https://anonymous.4open.science/r/GRGA-680F/README.md}{Anonymous GitHub}.
% Although traditional speech question answering (QA) has been well investigated, long-form speech meeting QA remains challenging due to intricate temporal dynamics, multi-speaker interactions and long-term semantic understanding. Existing retrieval-augmented generation (RAG) approaches which treat audio transcripts as flat text struggle with multi-hop reasoning, while current benchmarks lack diversity.
% To address this, we introduce \textbf{LongAudioQA}, a dataset feature a novel taxonomy that includes inferential, temporal, and acoustic-aware questions.
% We further propose a \textbf{Graph-based Retrieval-Generation Agent} (GRGA) modeled as a Partially Observable Markov Decision Process (POMDP). Mimicking human cognition, our agent iteratively \textbf{plans} query steps, \textbf{executes} graph-based tools (e.g., temporal filtering), and \textbf{reflects} to self-correct.
% Experiments show our agent outperforms standard RAG and long-context LLMs by \textbf{10\%} on complex tasks, offering superior interpretability and reduced hallucination.
\end{abstract}

\section{Introduction}
\renewcommand{\thefootnote}{*} 
\footnotetext{Corresponding Author: Dong Zhang}
% 数据集motivation：以往研究仅关注长会议的日志识别，或者做短对话的问答，还没有长会议的QA数据集，但是这个任务非常重要。所以我们构建这种数据集，并且从三个维度：semantic, speaker, time。
% 方法motivation 1、现有Speech LLM 也是文本为主、语音为辅 ,将语音映射到文本维度, 丢失信息；2、以往语音对话的QA都是短语音片段，他们提出的方法用于长语音QA，无法捕捉长期的记忆，而且以往研究完全忽略对话/会议中丰富的说话者特征（Role: Project Manager, Gender: Male, Stance: Conservative）。所以我们提出GRGA来解决以上问题，其中我们的多维度图可以构建文本、语音、说话者、时间等等信息，接着我们通过agent planning来检索我们的多维度图获取问题相关信息，最终生成答案，这样就可以有效解决前面两个挑战。
Long-form audio meeting understanding (LAMU) has attracted significant attention in speech processing. However, previous studies have only focused on transcription recognition for long-form multi-party meetings~\cite{alimeeting, alimeeting2,jain2024insectidentificationwildami}. Consequently, there is a lack of dedicated question answering (QA) datasets for long audio meetings, despite the critical importance of this task. To address this gap, we construct the \textbf{LongAudioQA} dataset for LAMU. Distinct from short-form conversation QA, it is designed to capture three core dimensions of long-form audio meeting content: complex semantics, multi-speaker interactions, and quite long timestamps.
 % (or question answering for short audio conversations)

\begin{figure}[t]
    \centering
    \includegraphics[width=1\linewidth]{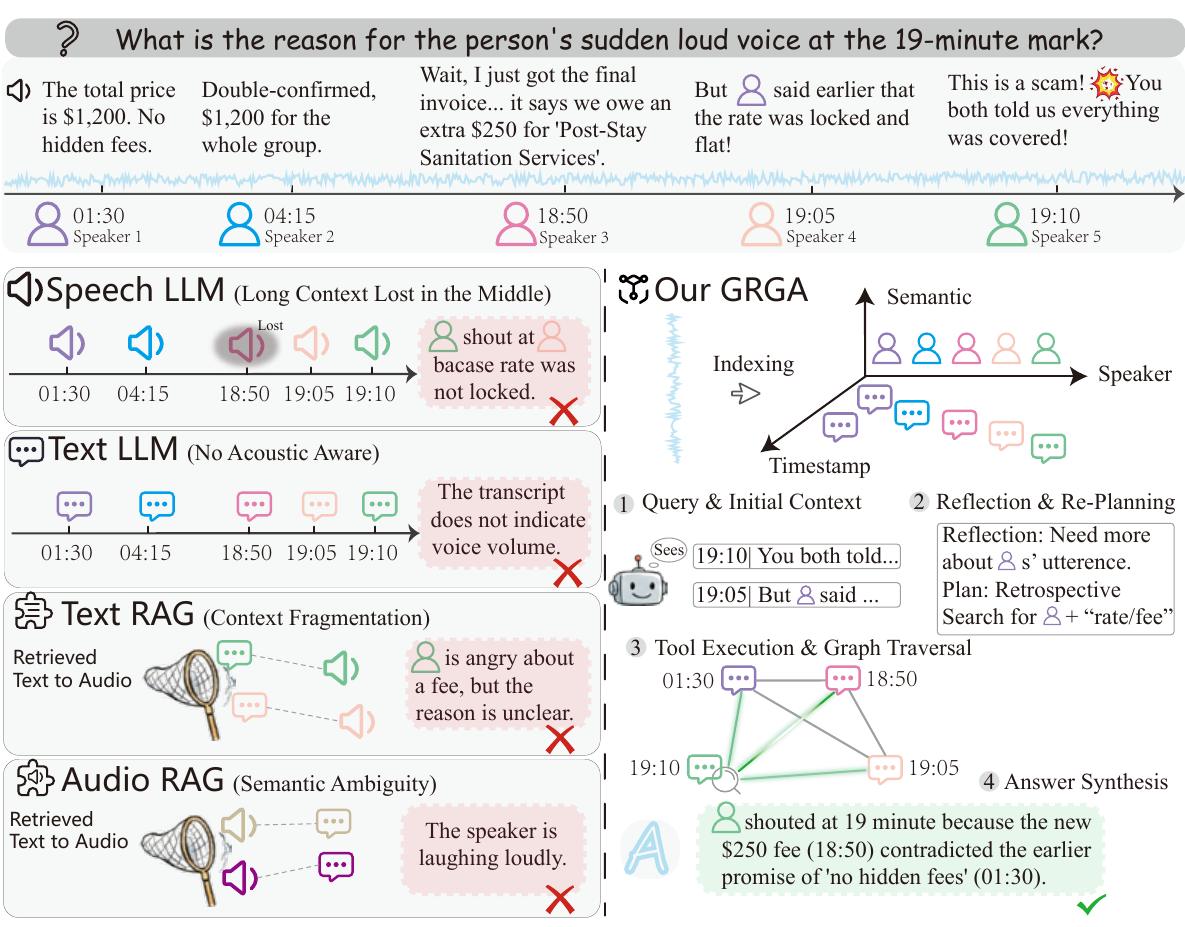}
\caption{\textbf{Existing Speech LLMs and our GRGA for long-form audio QA.} 
By indexing conversations across \textbf{Semantic}, \textbf{Speaker}, and \textbf{Timestamp} dimensions, our model enables precise reasoning.}
    \label{fig:intro}
\end{figure}

Although existing Speech LLMs~\cite{kimiteam2025kimiaudiotechnicalreport, Qwen3-Omni} have demonstrated strong performance across various tasks, they prioritize textual context by ASR over speech context~\cite{you-etal-2022-end,icassp/LinLCW0MLL24}. Specifically, these models typically map the input speech to the textual context, which inevitably leads to the loss of valuable acoustic information. 
For example, regarding the query ``sudden loud voice at the 19-minute mark'' in Figure \ref{fig:intro}, the inability to access voice volume prevents natural understanding and appropriate response (as the result of Text LLM in the Figure). We denote this phenomenon as \textbf{Acoustic Missing}.
Additionally, prior QA on speech conversations has been exclusively centered on short-form audio clips or dialogues ($<$ 30s) \cite{interspeech/JohnsonPSGGE24,tai/ZhaoJLWW25,longQA}. When applied to extremely long audio meetings, these methods fail to capture long-term dependencies. Let's return to the query in Figure \ref{fig:intro}, we not only need to locate the content of the 19-minute timestamp, but also to trace the answer back to the description of 1:30 (as our GRGA). Without an explicit design for long-range semantic understanding, even RAG approaches \cite{graphmpa,acl/dialogue-rag,icassp/WangSS00YS24} fail to identify the correct rationale (as Text RAG). We designate this phenomenon as \textbf{Context Fragmentation}.

To address these challenges, we propose the Graph-based Retrieval-Generation Agent (\textbf{GRGA}) model. Specifically, we first model heterogeneous features from the audio, including not only acoustic information (e.g., voice and tone) but also many speaker attributes (e.g., role: project manager, gender: male), into a unified multi-dimensional graph structure. Then, we leverage agent planning to retrieve question-relevant clues from this multi-dimensional graph and generate the final answer. In summary, we contribute:

$\bullet$ We design and construct the novel dataset \textbf{LongAudioQA} for LAMU. 

$\bullet$ We propose \textbf{GRGA} to handle both acoustic missing and context fragmentation.

$\bullet$ We conduct both automatic and human evaluation on three datasets of our LongAudioQA.

\label{sec:taxonomy}

\begin{table*}[t]
    \centering
    \renewcommand{\arraystretch}{1.6} 
    \small 
    
    \begin{tabularx}{\textwidth}{l l X}
        \toprule
        \textbf{Category} & \textbf{Core Competency} & \textbf{Subtasks \& Example Queries} \\
        \midrule
        
        % --- Factual ---
        % 使用自定义的 \categorylabel 命令
        \categorylabel{cfactual}{Factual} & Explicit Retrieval & 
        $\bullet$ \textit{Entity Retrieval:} ``Who mentioned the project code `Alpha'?'' \newline
        $\bullet$ \textit{Attribute Association:} ``What is the budget proposed by the Manager?'' \newline
        $\bullet$ \textit{Keyword Locating:} Matching specific text spans. \\
        \midrule
        
        % --- Inferential ---
        \categorylabel{cinferential}{Inferential} & Multi-hop Reasoning & 
        $\bullet$ \textit{Causal Inference:} Linking a problem raised early with the final decision. \newline
        $\bullet$ \textit{Coreference Resolution:} Identifying what ``that plan'' refers to. \newline
        $\bullet$ \textit{Stance Analysis:} Tracking how a speaker's attitude evolves. \\
        \midrule
        
        % --- Temporal ---
        \categorylabel{ctemporal}{Temporal} & Time Awareness & 
        $\bullet$ \textit{Absolute Localization:} ``What topic was discussed at the 30-min mark?'' \newline
        $\bullet$ \textit{Relative Sequencing:} ``What was discussed \textit{after} `market research'?'' \newline
        $\bullet$ \textit{Frequency Stats:} Counting term occurrences in a window. \\
        \midrule
        
        % --- Summarization ---
        \categorylabel{csummarization}{Summarization} & Aggregation & 
        $\bullet$ \textit{Topic Summarization:} Synthesizing consensus on ``backend architecture''. \newline
        $\bullet$ \textit{Speaker Profiling:} Summarizing a participant's main contributions. \\
        \midrule
        
        % --- Acoustic-Aware ---
        \categorylabel{cacoustic}{Acoustic-Aware} & Multi-modal Alignment & 
        $\bullet$ \textit{Emotion \& Intensity:} ``Who seemed most \textbf{agitated} when discussing the budget?'' \newline
        $\bullet$ \textit{Cross-modal Localization \& Causal:} ``What is the \textbf{reason} for the person's sudden loud voice at the 15-minute mark?'' \\
        
        \bottomrule
    \end{tabularx}
    
    \caption{\textbf{Taxonomy of Meeting Questions in our LongAudioQA.} Categories are distinguished by background colors: 
    \categorylabel{cfactual}{Factual}, 
    \categorylabel{cinferential}{Inferential}, 
    \categorylabel{ctemporal}{Temporal}, 
    \categorylabel{csummarization}{Summarization}, and 
    \categorylabel{cacoustic}{Acoustic-Aware}. 
    The \textbf{Acoustic-Aware} category uniquely requires grounding textual semantics with paralinguistic acoustic signals.}
    \label{tab:question_taxonomy}
\end{table*}

\section{Dataset Construction}

Existing speech QA benchmarks \cite{zhifei-etal-2025-audio} predominantly focus on short-context span extraction or simple intent classification. They often treat the dialogue as a flat sequence of text, neglecting the intricate graph-like structure of meeting interactions (e.g., speaker turns, cross-references, and temporal dynamics). Consequently, current models are rarely tested on their ability to perform multi-hop reasoning or temporal grounding over long-form recordings. For instance, answering a question like ``How did the speaker's attitude change after the 30-minute discussion?'' requires a model to not only localize information but also aggregate evidence across distinct time slices and model the causal dependencies between nodes. Therefore, we propose our \textbf{LongAudioQA}.

\subsection{Data Selection and Collection}
We mainly construct the speech question-answer pairs from the following raw dataset:

\textbf{AliMeeting.} 
AliMeeting~\cite{alimeeting, alimeeting2} is a large-scale Mandarin speech dataset tailored for multi-speaker meeting ASR and SD. A distinguishing feature of this corpus is the simultaneous recording of overlapping far-field audio and individual near-field references. This setup is designed to address the ``who said what when'' problem, testing the model's ability to handle speaker overlap and diarization in complex settings.

\textbf{AMI Meeting.} 
% The AMI Meeting Corpus~\cite{jain2024insectidentificationwildami} consists of meeting recordings captured in instrumented rooms with diverse acoustic properties, also designed for ASR. It utilizes synchronized close-talk and far-field microphones to capture interactions among primarily non-native English speakers. The dataset is characterized by its acoustic complexity, presenting significant challenges such as background noise and reverberation typical of real-world conference scenarios.
The AMI Conference Corpus~\cite{jain2024insectidentificationwildami} consists of meeting audio recordings captured using far-field microphones, primarily capturing interactions among non-native English speakers. It is characterized by acoustic complexity, presenting significant challenges commonly encountered in real conference settings, such as background noise and reverberation issues.

\textbf{DailyTalk.} 
% In contrast to the meeting-centric corpora aforementioned, DailyTalk~\cite{lee2022dailytalk} focuses on high-quality, open-domain dyadic conversations. It contains 2,541 dialogues derived from the DailyDialog dataset, inheriting its rich annotated attributes. Unlike the noisy conditions in AMI or AliMeeting, DailyTalk provides cleaner recordings with a focus on conversational flow and prosody, serving as a benchmark for standard interaction modeling. We concatenate 20 audio clips as a Short-context Control Group to make sure that our method works well on long audio segments while maintaining its effectiveness on short audio segments as well.
In contrast to the aforementioned conference-centric corpora, DailyTalk~\cite{lee2022dailytalk} focuses on high-quality open-domain dyadic conversations. This corpus comprises 2,541 dialogues. DailyTalk provides clean acoustic environment with an emphasis on conversational fluency and prosodic features. We concatenated 20 clips to create inputs of intermediate length ($<$10 minutes). While shorter than full meetings, this duration serves to ensure our method performs well on longer audio segments while maintaining effectiveness on shorter ones.

\subsection{Question Definition and Data Annotation}
% To bridge the above gap, we construct a novel dataset (Table \ref{tab:question_taxonomy}) designed to push the boundaries of long-form meeting comprehension. Unlike previous works that rely solely on factual retrieval, our dataset introduces a diverse taxonomy of questions—ranging from Factual and Inferential to Temporal and Acoustic-aware. This hierarchical design allows us to systematically evaluate a model's capability to transition from explicit pattern matching to high-level semantic reasoning within a complex interaction graph.
To bridge the gap, we constructed a novel dataset designed to expand the boundaries of long-form meeting understanding. Unlike existing research that relies solely on factual retrieval, our dataset introduces a diversified question taxonomy, encompassing factual, inferential, temporal, and acoustic-aware as shown in Table \ref{tab:question_taxonomy}). This design enables us to systematically evaluate models' ability to transition from explicit pattern matching to advanced semantic reasoning within complex interaction graphs.
Following methodologies for automated data annotation using LLMs ~\cite{icml/00040C0SRC00PY025,icml/00040S0CG0CZY0025}, we employ a hybrid strategy combining large language model-driven automated annotation with human-assisted verification. Specifically, this approach leverages large language models to generate questions from specific local segments of raw conference data based on predefined query types and examples. The models then search conference content for answers and extract detailed evidence. This process is iterative: each batch of generated question-answer pairs undergoes human verification to ensure quality before advancing to the next iteration. While the generation process leverages an “oracle” mechanism to access localized evidence, the core challenge of this study lies in retrieving answers from global, long-context conference data without prior knowledge of relevant segments. This evaluates the model's ability to overcome context fragmentation.

\subsection{Dataset Quality Control}

\begin{figure}[h!]
    \centering
    \includegraphics[width=0.8\linewidth]{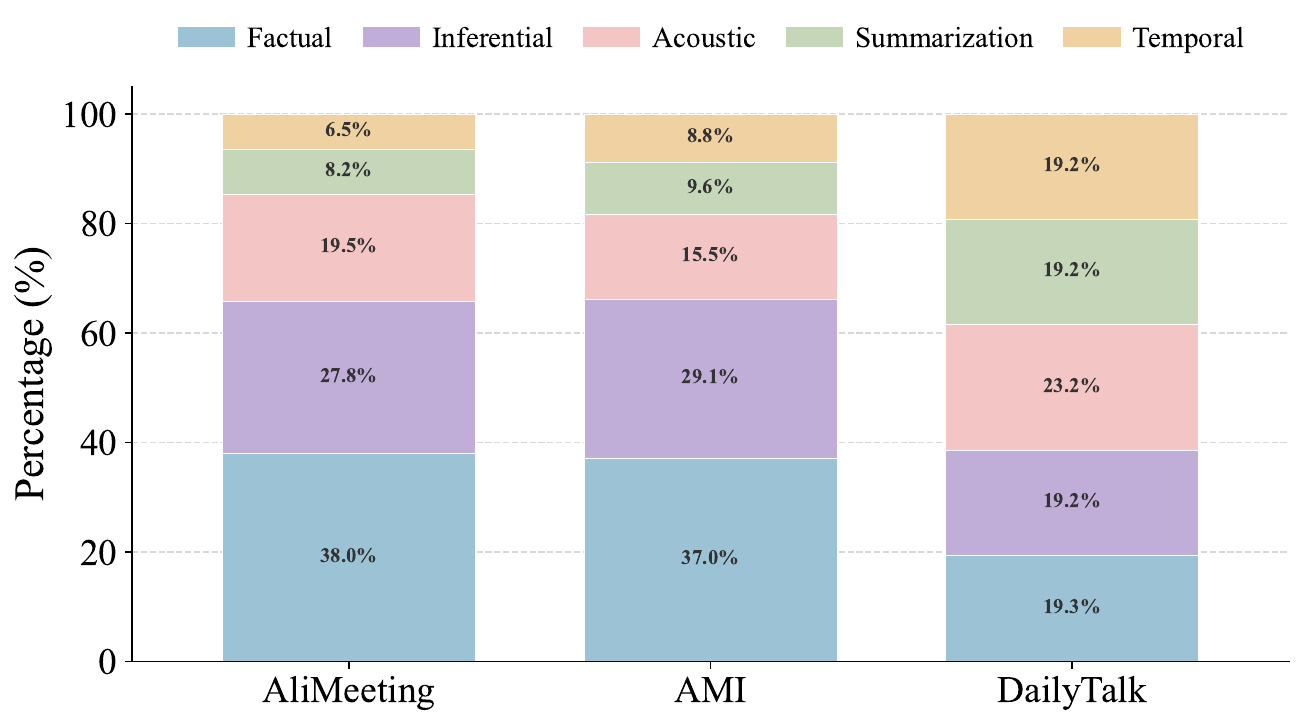}
    \caption{Question Distribution Across Corpora}
    \label{fig:qas}
\end{figure}

Given that automated generation inevitably introduces noise, we enforce a rigorous verification protocol to ensure data reliability. Expert annotators inspect candidate samples to discard hallucinated questions (where answers are absent), rewrite ambiguous references, and validate logic depth (hop count) to ensure complexity balance. Specifically, we require that the generated timestamp evidence yields an Intersection-over-Union (IoU) of \( > 0.9 \) with the ground truth. This process results in a high-quality dataset with a human inter-annotator agreement rate of \( \kappa = 0.91 \) (Cohen's Kappa). Finally, the duration and sample statistics of our LongAudioQA, along with the question distribution, are summarized in Table~\ref{tab:dataset_stats} and Figure~\ref{fig:qas}.

\begin{table}[t]
    \centering
    \resizebox{\linewidth}{!}{ 
    \begin{tabular}{lrrrrr}
        \toprule
        \textbf{Dataset} & \textbf{Dur (h)} & \textbf{Avg. Dur (s)} & \textbf{Avg. Turns} & \textbf{\# Q\&A} & \textbf{\# Dial.} \\
        \midrule
        AliMeeting & 14.91 & 1,935 & 815 & 3,013  & 28  \\
        AMI         & 18.24 & 1,930 & 757 & 3,243  & 34  \\
        DailyTalk   & 21.59 & 610   & 186 & 10,200 & 128 \\
        \bottomrule
    \end{tabular}  
    }
    \caption{Statistical information among DailyTalk, AMI, and AliMeeting datasets. \textbf{Dur}: Duration, \textbf{Avg}: Average.}
    \label{tab:dataset_stats}
\end{table}

% 饼图

\section{Methodology}
\begin{figure*}
    \centering
    \includegraphics[width=1\linewidth]{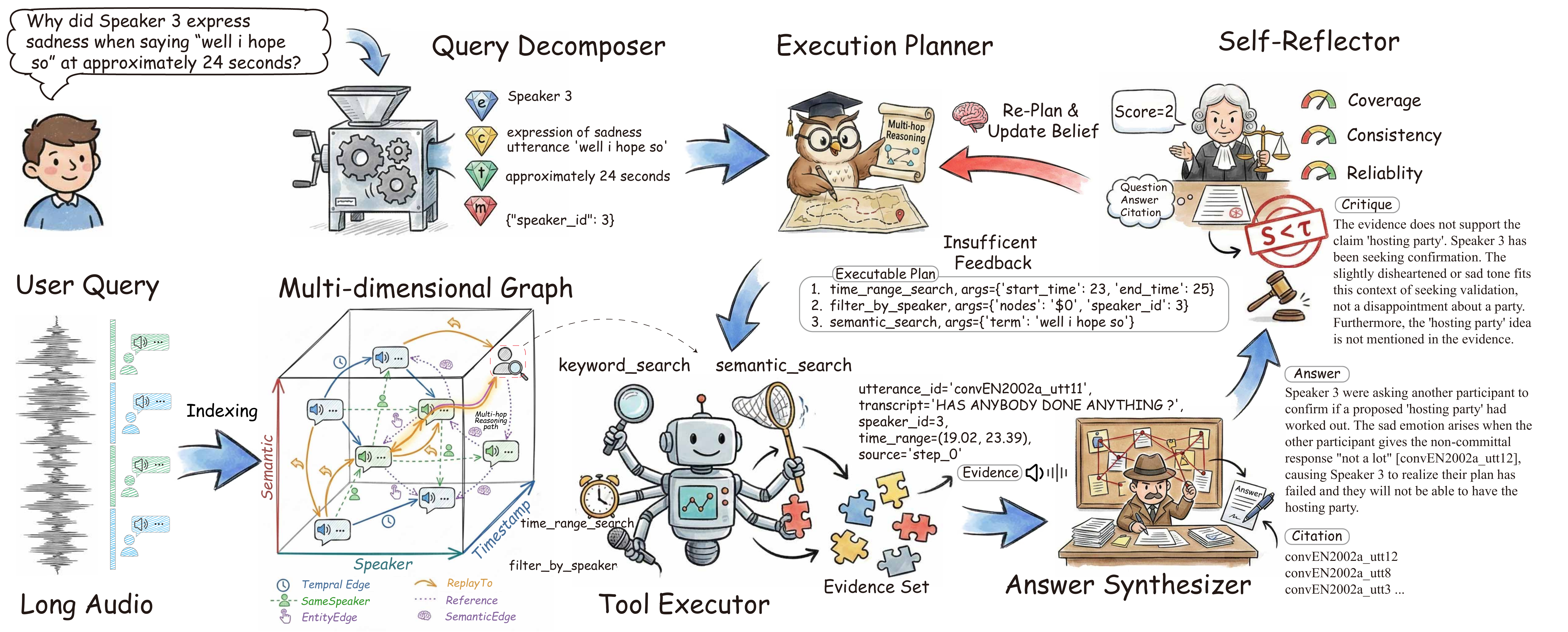}
    \caption{The overall architecture of our graph-based retrieval-generation agent \textbf{GRGA}. 
The framework models long-form audio as a \textbf{Multi-dimensional Graph} (bottom-left) to capture semantic, temporal, and speaker dependencies. 
Our GRGA Planning Process consists of: 
\textbf{Query Decomposition}, \textbf{Planning}, \textbf{Execution}, \textbf{Synthesis}, and \textbf{Reflection}.}
\label{fig:method}
\end{figure*}
\label{sec:method}

The core challenge in long-form audio QA lies in handling composite queries that involve semantic, temporal, and speaker relationships. Existing methods often rely on ``one-shot'' retrieval, which fails to capture the intricate dependencies in meeting data. To address this, we propose a novel Graph-based Retrieval-Generation Agent (\textbf{GRGA}) that mimics the cognitive process of a human expert: it does not merely search, but rather \textit{plans} and \textit{reflects}.

\textbf{Cognitive Inspiration.}
To design an effective agent, we draw inspiration from the cognitive strategies human experts employ to navigate complex meetings. We observe that resolving composite queries typically adheres to a \textbf{``Search-Reason-Verify''} cognitive loop, which adapts dynamically to the query type. For instance, in \textit{factual} tasks (e.g., verifying budget details), humans execute a targeted keyword scan followed by contextual verification—a mechanism we emulate through retrieval tools. Conversely, \textit{inferential} tasks (e.g., discerning the rationale behind a rejection) necessitates maintaining working memory while traversing causal chains. Similarly, humans naturally construct a \textit{mental timeline} to address \textit{temporal queries}, whereas \textit{summarization} involves synthesizing fragmented details into high-level concepts. Consequently, we architect our agent to replicate these cognitive behaviors by incorporating a \textbf{Planner} for logical orchestration, a \textbf{Synthesizer} module for holistic information aggregation, and a \textbf{Self-Reflector} for answer verification.

\subsection{Problem Formulation}
\label{sec:pomdp_formulation}

We formulate the task as a \textbf{Partially Observable Markov Decision Process (POMDP)}~\cite{POMDPs}, defined by the tuple $\mathcal{M} = \langle \mathcal{S}, \mathcal{A}, \mathcal{T}, \mathcal{R}, \Omega, \mathcal{O}, \gamma \rangle$. The pre-constructed multi-dimensional graph $\mathcal{G}$ serves as the environment with which the agent interacts.  

\textbf{Belief State ($b_t$) \& Policy ($\pi$)}: The belief state $b_t$, represented by the query $Q$ and interaction history $H_t$, serves as the sufficient statistic for decision-making. We leverage a pre-trained LLM as the policy $\pi(a_t \mid b_t)$. Implemented via the \textit{Query Decomposer} and \textit{Execution Planner}, the policy utilizes in-context reasoning to determine the next optimal action without parameter updates.

\textbf{Action ($\mathcal{A}$) \& Observation ($\Omega$)}: The action space $\mathcal{A}$ comprises topological retrieval operations. The \textit{Tool Executor} executes $a_t$ on $\mathcal{G}$, yielding a partial observation $o_t \in \Omega$ (e.g., specific acoustic cues or semantic nodes), which updates $b_t$.

\textbf{Reward ($\mathcal{R}$)}: The \textit{Self-Reflector} acts as a verifier, providing a heuristic reward $r_t$. It evaluates whether $o_t$ aligns with the reasoning chain, prompting the planner to prune incorrect paths or terminate generation.

Since the model is training-free, our objective is to approximate the optimal trajectory $\tau^* = \{b_0, a_0, \dots, a_T\}$ that maximizes the reliability of the final answer. The process terminates when the \textbf{Answer Synthesizer} determines that sufficient evidence is gathered ($r_t > \text{threshold}$) or the maximum step limit is reached. Figure \ref{sec:method} illustrates our framework.

\subsection{Audio-to-Graph Indexing}

\subsubsection{Acoustic Semantic Alignment}
To initialize the graph nodes, we transform the continuous audio stream into discrete, semantically meaningful units with precise temporal boundaries.

\textbf{VAD and Transcription.} We employ FSMN-VAD to segment audio $\mathcal{A}$ into clips, filtering silence. These clips are transcribed by a high-fidelity ASR model to obtain text $T$. We apply CTC-based Forced Alignment: for each token $w_i$, we perform constrained Viterbi decoding to align text with acoustic features, yielding exact boundaries $[t_{start}, t_{end}]$. 

\textbf{Speaker Diarization.}
Identifying ``who spoke'' is insufficient; understanding ``who they are'' (e.g., role, stance) is key to reasoning. We propose a multimodal profiling mechanism.
We extract speaker embeddings using ERes2NetV2 and perform incremental clustering. If the cosine similarity between a new embedding and existing clusters is below a threshold $\eta=0.8$, a new speaker ID is created.

\textbf{Semantic Role Generation.} 
A raw ID (e.g., \texttt{Speaker\_0}) lacks semantic context. We construct a \textbf{Speaker Profile} $\mathcal{P}_k$ for each unique speaker $S_k$. We aggregate all utterances belonging to $S_k$ and prompt an LLM to distill attributes:
\begin{equation}
    \mathcal{P}_k = \text{LLM}(\text{Concat}(\{v_i.\text{text} \mid v_i.\text{spk} = S_k\}))
\end{equation}
The output is a structured profile, e.g., \texttt{\{Role: Project Manager, Gender: Male, Stance: Conservative\}}. This profile is stored as a node attribute.

\subsubsection{Multi-dimensional Graph Modeling}

Beyond the raw meeting transcripts, we also incorporate the acoustic features (e.g., speaker identity, start/end timestamps, and corresponding speech) and explicitly model the dialogue flow as a heterogeneous graph  $\mathcal{G} = (\mathcal{V}, \mathcal{E}) $. 

\textbf{Nodes ($ \mathcal{V} $):} Each utterance is treated as a fundamental node $ v_i $, enriched with attributes including text transcripts, speaker identity, start/end timestamps, and corresponding speech, etc.

\textbf{Edges ($ \mathcal{E} $):} To support the diverse reasoning tasks, we construct multiple edge types: 

(1) \textit{Temporal Edges ($e_{temp}$)} connecting adjacent utterances ($ v_i \rightarrow v_{i+1} $); 

(2) \textit{Reply-To Edges ($e_{reply}$):} Captures conversational turns. We establish a directed edge $v_i \rightarrow v_j$ if $v_j$ occurs within a dynamic window $\Delta t < 5s$ after $v_i$ and speakers are different; 

(3) \textit{SameSpeaker Edges ($e_{spk}$)} Connects nodes $v_i, v_j$ where $\text{spk}(v_i) = \text{spk}(v_j)$. This enables the agent to aggregate scattered opinions of a specific person;

(4) \textit{Entity Edges ($e_{ent}$)} derived from entity keyword co-occurrence or discourse relations.

(5) \textit{Semantic Edges ($e_{sem}$)} To solve context forgetting, we employ an LLM to detect coreference chains. If $v_j$ contains pronouns (e.g., ``that idea'') referring to an entity in $v_i$, we create a semantic link $v_j \xrightarrow{sem} v_i$.

\subsection{GRGA Planning Process}

\paragraph{Belief Initialization: Query Decomposition.}

Given a user query $Q$, the belief state is initialized by projecting the unstructured query into a structured constraint space $\mathcal{C}$. We define the decomposition function $f_{\text{dec}}: \mathcal{Q} \to \mathcal{C}$ as:
\begin{equation}
    \mathcal{C} = f_{\text{dec}}(Q) = \{ c_e, c_c, c_t, c_m \}
\end{equation}
where $c_e, c_c, c_t, c_m$ denote constraints on \textbf{entities}, \textbf{concepts}, \textbf{time}, and \textbf{metadata}, respectively. The initial belief is set as $b_0 = H_0 = \{Q, \mathcal{C}\}$.

\paragraph{Policy Network: Execution Planning.}
The Execution Planner approximates the policy $\pi_\theta$ parameterized by an In-Context Learning (ICL )~\cite{dong-etal-2024-survey-icl} guided LLM. At step $t$, it generates a multi-step plan $\mathcal{P}_t$ conditioned on the current belief $b_t$:
\begin{equation}
    \mathcal{P}_t \sim \pi_\theta(\cdot \mid b_t)
\end{equation}
The plan $\mathcal{P}_t$ is a sequence of atomic operations derived from the Action Space $\mathcal{A}$ (see Table~\ref{tab:tool_definitions}):
\begin{equation}
    \mathcal{P}_t = [op_1, op_2, \dots, op_k], \quad op_i \in \mathcal{A}
\end{equation}
This formulation enables \textbf{long-horizon planning}, allowing the agent to chain logical operations (e.g., \texttt{Search} $\to$ \texttt{Filter}) before interacting with the environment.

\paragraph{Environment Interaction: Tool Execution.}
The Execution Engine functions as the transition interface. It executes the plan $\mathcal{P}_t$ on the graph $\mathcal{G}$ to obtain an observation $o_t$:
\begin{equation}
    o_t = \text{Exec}(\mathcal{P}_t, \mathcal{G}) = \{s_1, s_2, \dots, s_n\}
\end{equation}
where each segment $s_i = (\text{text}_i, \text{time}_i, \text{spk}_i)$.
Upon receiving $o_t$, the agent updates its belief state via a deterministic transition function $\psi$:
\begin{equation}
    b_{t+1} = \psi(b_t, \mathcal{P}_t, o_t) = b_t \oplus \{\mathcal{P}_t, o_t\}
\end{equation}

\paragraph{Reward Estimation: Synthesis \& Reflection.}
To guide the reasoning trajectory refinement, we employ a two-stage mechanism as reward function.

1) Answer Synthesis.
The synthesizer generates a candidate answer $\hat{A}_t$ and citations $\text{Cite}_t$ based on the accumulated evidence in $b_{t+1}$:
\begin{equation}
    (\hat{A}_t, \text{Cite}_t) = f_{\text{syn}}(Q, b_{t+1})
\end{equation}

2) Reflection as Sparse Reward.
The Reflector evaluates the logical entailment between the evidence $E \subset b_{t+1}$ and the answer $\hat{A}_t$, assigning a verification score $s_{\text{ver}} \in [0, 1]$:
\begin{equation}
    s_{\text{ver}} = f_{\text{ref}}(Q, \hat{A}_t, E)
\end{equation}
The reward $r_t$ is defined as a threshold function:
\begin{equation}
    r_t = \begin{cases} 
    1 & \text{if } s_{\text{ver}} \ge \tau \quad (\text{Success, Terminate}) \\
    -\beta & \text{if } s_{\text{ver}} < \tau \quad (\text{Failure, Re-plan})
    \end{cases}
\end{equation}
If $r_t < 0$, the negative feedback $fb_t$ (critique) is injected into the belief state: $b_{t+1} \leftarrow b_{t+1} \cup \{fb_t\}$, prompting the policy $\pi$ to generate a corrective plan $\mathcal{P}_{t+1}$ in the next iteration.

\section{Experimentation}
\begin{table*}[t] 
    \centering
    \renewcommand{\arraystretch}{1.2}
    \setlength{\tabcolsep}{4pt}
    \resizebox{\textwidth}{!}{%
        \begin{tabular}{l  ccccc  ccccc  ccccc}
            \toprule
            \multirow{2.5}{*}{\textbf{Method}} & \multicolumn{5}{c}{\textbf{AliMeeting }} & \multicolumn{5}{c}{\textbf{AMI Meeting}} & \multicolumn{5}{c}{\textbf{DailyTalk}} \\
            \cmidrule(lr){2-6} \cmidrule(lr){7-11} \cmidrule(lr){12-16}
             & \textbf{Fact.} & \textbf{Infer.} & \textbf{Temp.} & \textbf{Summ.} & \textbf{Acou.}  & \textbf{Fact.} & \textbf{Infer.} & \textbf{Temp.} & \textbf{Summ.} & \textbf{Acou.} & \textbf{Fact.} & \textbf{Infer.} & \textbf{Temp.} & \textbf{Summ.} & \textbf{Acou.} \\
            \midrule
            
            % Group 1: End-to-End Speech LLM
            \multicolumn{14}{l}{\textit{\textbf{Speech as Context}}} \\
            Qwen3-Omni          & 29.31 & 27.33 & 15.81 & 19.06  & 16.50 & 20.96 & 26.38 & 15.12 & 16.67 & 12.15 & 34.26 & 38.43 & 2.43 &  2.11 & 9.39  \\
            Audio Flamingo 3    & 35.70 & 35.23 & 13.19 & 18.09 & 32.89 & 16.04 & 27.94 & 12.62 & 10.24 & 17.06 & 65.13 & 61.31 & 3.83 & 2.44 & 40.09  \\
            MiMo-Audio          & 54.50 & 54.50 & 16.92 & 39.91 & 25.15 & 52.75 & 64.74 & 35.34 & 38.30 & 17.17 & 81.57 & 78.65 & 5.58 & 3.46 & 31.44  \\
            
            % Group 2: Long-Context Text Baseline
            \midrule
            \multicolumn{14}{l}{\textit{\textbf{Transcription as Context}}} \\
            Qwen3-Omni		 & 47.66 & 63.53 & 38.34 & 32.65 & 11.70 & 53.55 & 46.66 & 37.32 & 43.63 & 14.13 & 80.11 & 77.14 & 62.71 & 15.71 & 18.27 \\
            Audio Flamingo 3 & 32.12 & 45.16 & 28.23 & 21.06 &  9.69 & 48.52 & 40.21 & 22.12 & 29.16 & 12.36 & 76.43 &  75.26 & 58.32 & 13.42 & 15.85 \\
            MiMo-Audio		 & 48.02 & 64.62 & 38.29 & 29.43 & 15.72 & 53.62 & 44.10 & 37.14 & 41.49 & 17.94 & 80.67 &  76.16 & 62.46 & 15.35 & 17.52 \\
            
            % Group 3: Text RAG Baseline
            \midrule
            \multicolumn{14}{l}{\textit{\textbf{Both Speech and Transcription as Context with RAG}}} \\
            BGE-M3       & 43.24 & 29.68 & 31.56 & 15.43 & 18.45 & 32.56 & 24.56 & 23.78 & 25.36 & 25.34 & 46.72 & 36.31 & 32.46 & 34.79 & 26.21  \\
            % Audio Flamingo 3 & 52.86 & 37.69 & 31.75 & 22.57 & 36.2 & 34.74 & 48.91 & 25.80 & 26.41 & 25.50 & 64.11 &  68.24 & 46.71 & 35.25 & 48.46 \\
            % MiMo-Audio       & 54.29 & 45.39 & 32.05 & 38.28 & 37.43 & 46.18 & 53.56 & 33.95 & 34.47 & 27.20 & 79.54 & 76.54 & 48.25 & 38.13 & 43.28  \\
            
            % Group 3: Audio RAG Baseline
            % \midrule
            % \multicolumn{14}{l}{\textit{\textbf{Audio  Retrieval-Augmented Generation}}} \\
            CLASP      & 25.37 & 19.54 & 18.92 & 14.76 & 16.32 & 24.12 & 20.31 & 22.76 & 11.60 & 14.79 & 30.58 & 33.42 & 27.34 & 13.52 & 11.81 \\
            % Audio Flamingo 3 & 52.86 & 37.69 & 31.75 & 22.57 & 36.2 & 34.74 & 48.91 & 25.80 & 26.41 & 25.50 & 64.11 &  68.24 & 46.71 & 35.25 & 48.46 \\
            % MiMo-Audio       & 54.29 & 45.39 & 32.05 & 38.28 & 37.43 & 46.18 & 53.56 & 33.95 & 34.47 & 27.20 & 79.54 & 76.54 & 48.25 & 38.13 & 43.28  \\       
            
            % Group 4: OURS (Highlighted)
            \cdashline{1-16}
            % \multicolumn{14}{l}{\textit{\textbf{(Our) GRGA}}} \\
            \rowcolor{rowgray} % 背景高亮
            GRGA(ours) &  \textbf{57.31} & \textbf{66.45} & \textbf{39.48} & \textbf{44.29} & \textbf{39.91} & \textbf{59.46} & \textbf{65.29} & \textbf{38.56} & \textbf{48.46} & \textbf{35.68} & \textbf{85.25} & \textbf{79.06} & \textbf{63.58} & \textbf{61.10} & \textbf{52.32}\\
            \bottomrule
        \end{tabular}%
    }
    \caption{\textbf{Main results on accuracy (\%).} We compare our proposed method against End-to-End Speech LLM and standard RAG baselines across three datasets. \textbf{Bold} indicates the best performance. \textit{Fact.}: Factual, \textit{Infer.}: Inferential, \textit{Temp.}: Temporal, \textit{Summ.}: Summarization, \textit{Acou.}: Acoustic reasoning.}
    \label{tab:main_results}
\end{table*}

% \subsection{Datasets}
% We conduct experiments on three distinct datasets to evaluate our model across different acoustic environments and speaking styles.
To validate the effectiveness of our GRGA, we conduct extensive experiments on our LongAudioQA dataset.

\subsection{Baselines and Implementation}

First, we compare three recent most competitive multi-modal LLMs: Qwen3-Omni~\cite{Qwen3-Omni}, \textbf{Audio Flamingo 3}~\cite{goel2025audioflamingo3advancing}, and MiMo-Audio ~\cite{coreteam2025mimoaudio}, which are considered as SOTAs for speech or spoken text QA. Therefore, we implement them in two forms: 1) use whole meeting \textbf{Speech as Context} and 2) use whole meeting \textbf{Transcription as Context}.

Second, we investigate two tailored RAG approaches for audio QA: \textbf{BGE-M3}~\cite{chen2024bgem3embeddingmultilingualmultifunctionality} retrieving meeting text through a query, then merging the retrieved text and corresponding audio into the model (TextRAG) and \textbf{CLASP}~\cite{CLASP}, which are also considered as the SOTAs for RAG-based audio QA, retrieving meeting audio through a query, then merging the retrieved audio and corresponding text into the model (AudioRAG). They cast both meeting \textbf{Speech and Transcription as Context}. Notably, both of them adopt the same LLMs as ours to implement. More details of our GRGA and above baselines can refer to Appendix~\ref{experimental_details}. 

\subsection{Evaluation Metric}

To rigorously assess the reasoning capabilities of our model, we report \textbf{Semantic Accuracy} across all datasets. 
Traditional n-gram metrics (e.g., BLEU, Exact Match) often fail to capture the true validity of generated responses, as they penalize correct answers that differ in lexical surface forms from the ground truth. 
Drawing upon recent methodologies in complex reasoning evaluation \cite{mishra-etal-2025-investigating, shui-etal-2023-comprehensive}, we move beyond rigid string matching and establish a robust, automated evaluation pipeline that prioritizes semantic equivalence and logical soundness.

Specifically, we employ a high-capability LLM (LLM-as-a-Judge) to approximate human-level judgment. 
Formally, let $\mathcal{D} = \{(q_i, a_i)\}_{i=1}^{N}$ denote the evaluation dataset, where $q_i$ is the query and $a_i$ is the ground-truth answer. Let $\hat{a}_i$ represent the model-generated response. 
We define a semantic judgment function $\mathcal{J}$, parameterized by an external expert model (e.g., GPT-OSS-120B~\cite{openai2025gptoss120bgptoss20bmodel}):
\begin{equation}
    s_i = \mathcal{J}(q_i, a_i, \hat{a}_i) \in \{0, 1\},
\end{equation}
where $s_i = 1$ if and only if $\mathcal{J}$ determines that $\hat{a}_i$ entails the same semantic information as $a_i$, and $0$ otherwise. 
The judge is prompted to disregard stylistic differences and focus solely on factual consistency and the correctness of reasoning. 
Finally, the Semantic Accuracy is computed as the expectation of correct judgments:
\begin{equation}
\text{Accuracy} = \frac{1}{N} \sum_{i=1}^{N} s_i \times 100\%.
\end{equation}

This approach ensures a fair comparison by validating whether the model successfully retrieves and reasons over the core knowledge, regardless of its output phrasing.

\subsection{Main Results}
Table~\ref{tab:main_results} shows the performance comparison of all baselines of our GRGA on our proposed three datasets for Long-form speech meeting understanding. From this table, we can see:

\textbf{Overcoming Context Limitations.}
 End-to-End Speech LLMs (e.g., Qwen3-Omni) degrade sharply on long-form datasets, with AudioFlamingo3 dropping from $ \sim 65\% $ on DailyTalk to $ \sim 16\% $ on AMI. This confirms that fixed context windows hinder reasoning in long-form audio. Conversely, our GRGA maintains robust performance, outperforming the strongest baseline (MiMo-Audio) on AMI, validating the scalability of our graph-based retrieval beyond context limits.

\textbf{Planning vs. Naive Retrieval.}
Comparisons with RAG baselines highlight the failure of ``one-shot'' retrieval. Text RAG, while effective for factual questions, struggles significantly with inferential questions ($ 24.6\% $ vs. ours $ 65.3\% $ on AMI). This proves that vector similarity alone cannot capture multi-hop dependencies. Our Query Planner bridges this gap by decomposing queries into logical chains. Additionally, the poor performance of Audio RAG ($ \sim20\% $) indicates that raw acoustic retrieval is too noisy, justifying our use of a structured graph intermediate.

\begin{table}[t]
    \centering
    \renewcommand{\arraystretch}{1.15}
    \resizebox{\columnwidth}{!}{
        \begin{tabular}{l c c c}
            \toprule
            \multirow{2}{*}{\textbf{Dataset}} & \multirow{2}{*}{\textbf{WER} ($\downarrow$)} & \multicolumn{2}{c}{\textbf{DER} ($\downarrow$)} \\
            \cmidrule(lr){3-4}
             & & \textbf{Collar = 0 s} & \textbf{Collar = 0.25 s} \\
            \midrule
            DailyTalk       & 0.85  & 6.42  & 3.41  \\
            AliMeeting-far  & 20.38 & 16.80 & 13.10 \\
            AMI-sdm         & 18.84 & 17.60 & 14.90 \\
            \bottomrule
        \end{tabular}
    } 
    \caption{\textbf{Error analysis on benchmark datasets.} We report Word Error Rate (WER) and Diarization Error Rate (DER) under strict ($ 0\,\text{s} $) and standard ($ 0.25\,\text{s} $) collar (tolerance
) settings. All metrics are reported in percentage (\%), and $\downarrow$ denotes that lower values are better.}
    \label{tab:upstream_errors}
\end{table}

\subsection{Analysis and Discussion}

\begin{figure}[t]
    \centering
    \includegraphics[width=\linewidth]{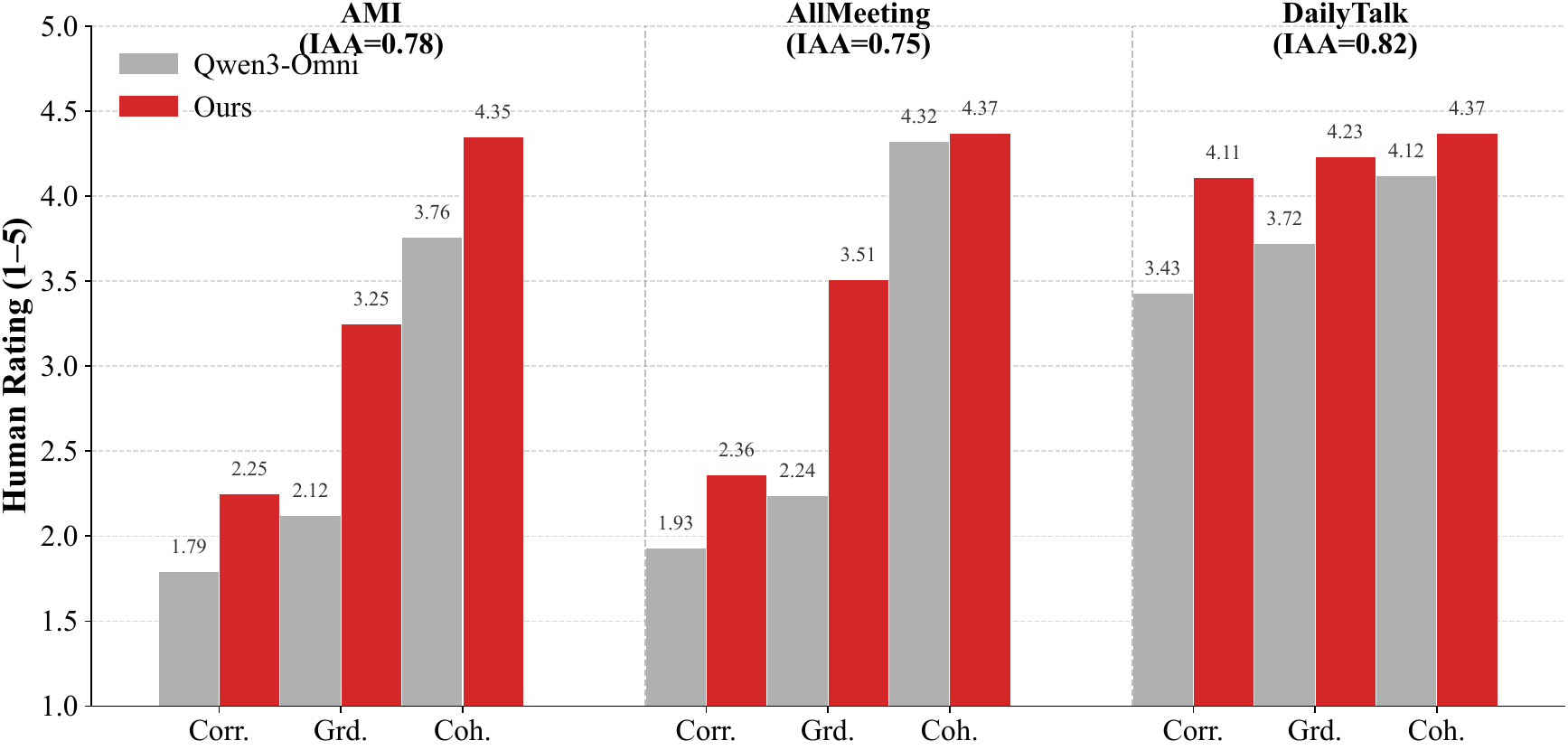} 
    \caption{\textbf{Human evaluation results.} Our method shows significant improvements, particularly in Groundedness across complex meeting scenarios.}
    \label{fig:human_eval}
\end{figure}

\textbf{Human Evaluation.} 
As illustrated in Figure~\ref{fig:human_eval}, our method consistently outperforms Qwen3-Omni across all datasets and metrics. Notably, the most substantial gap appears in \textbf{Groundedness} (e.g., \textbf{+1.27} on AllMeeting). While the baseline generates fluent text (high Coherence), it frequently lacks evidentiary support in multi-party dialogues. Our superior grounding directly translates to higher \textit{Correctness}, confirming that precise temporal anchoring effectively reduces factual errors.

\textbf{Upstream Performance Analysis.}
Table~\ref{tab:upstream_errors} highlights the noise disparity across datasets. DailyTalk serves as a clean baseline with minimal errors (WER $ 0.85\% $). Conversely, AMI and AliMeeting present high noise rates, with WERs $\sim 20\% $ and DERs $ > 13\% $. Despite these significant upstream errors in transcription and diarization, our GRGA maintains high QA accuracy (Table~\ref{tab:main_results}), demonstrating the robustness of our iterative planning and reflection mechanisms against noisy inputs.

\begin{figure}
    \centering
    \includegraphics[width=1\linewidth]{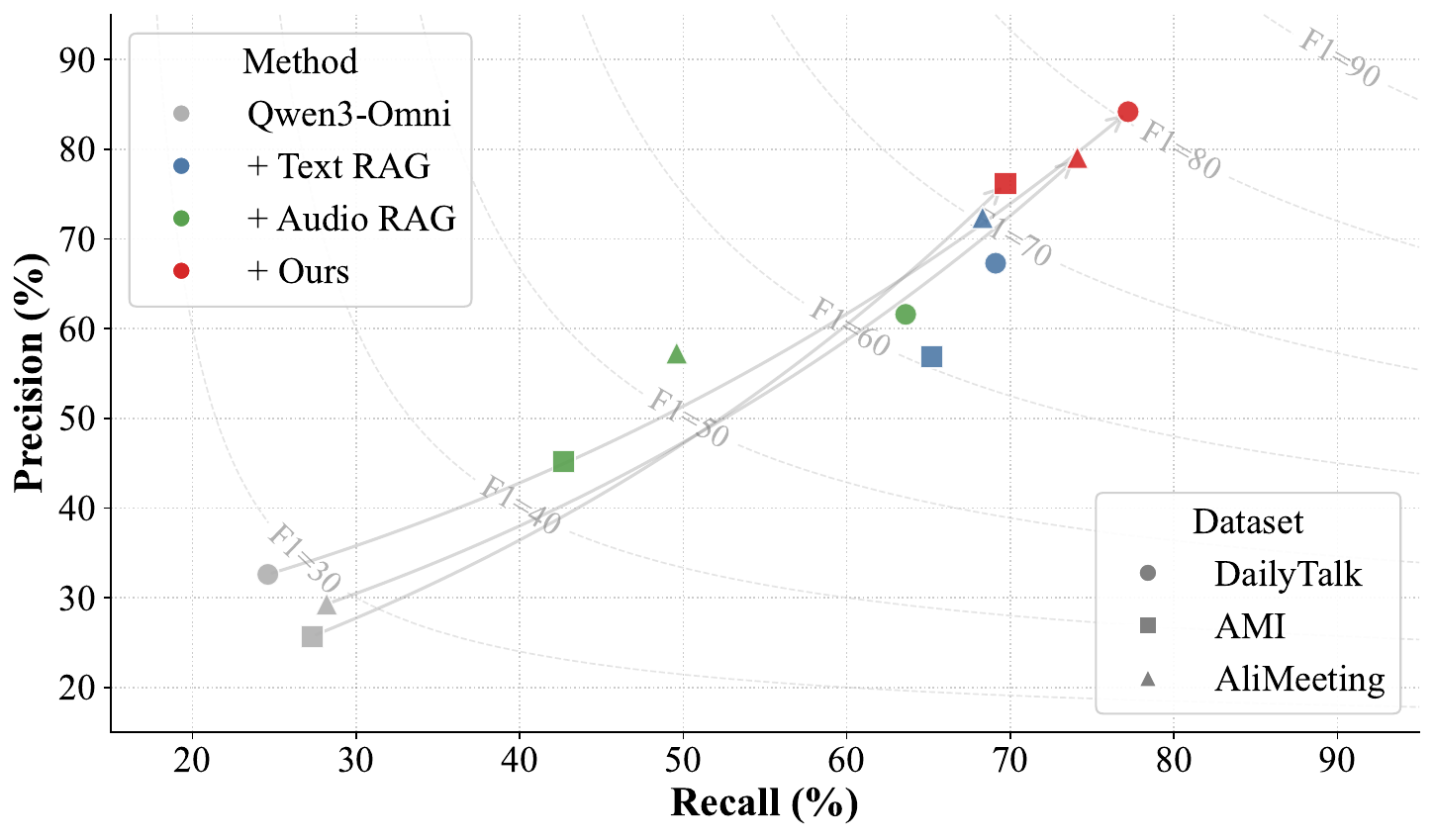}
    \caption{\textbf{Citation Precision-Recall Trade-off}. Results are reported in \% with a $\pm 2$s tolerance.}
    \label{fig:pr_scatter}
\end{figure}

\textbf{Evidence Recall \& Precision Analysis.}
To assess the model's ability to locate supporting evidence, we report the Precision (P) and Recall (R) of generated citations (timestamps) against ground truth spans (with a $\pm 2$s tolerance) in Figure~\ref{fig:pr_scatter} and Table~\ref{tab:pr_tab}. 
Our GRGA significantly outperforms baselines across all datasets. Specifically, compared to standard Text RAG, our method improves Precision by $ +19.3\% $ on AMI and $ +6.7\% $ on AliMeeting. 
This indicates that our \textbf{Query Planner} effectively filters out irrelevant noise, while the \textbf{Reflection} mechanism ensures that retrieved segments are strictly relevant to the answer, minimizing hallucinated citations common in naive retrieval approaches.

\begin{table}[t]
    \centering
    \renewcommand{\arraystretch}{1.2}
    \setlength{\tabcolsep}{3.5pt}
    
    \resizebox{\columnwidth}{!}{%
        \begin{tabular}{l | c c | c c c c c}
            \toprule
            \multirow{2}{*}{\textbf{Ablation Settings}} & \multicolumn{2}{c|}{\textbf{Overall}} & \multicolumn{5}{c}{\textbf{AMI }} \\
            \cmidrule(lr){2-3} \cmidrule(lr){4-8}
             & \textbf{Avg.} & \textbf{$\Delta$} & \textbf{Fact.} & \textbf{Infer.} & \textbf{Temp.} & \textbf{Summ.} & \textbf{Acou.} \\
            \midrule
            
            % --- Full Model (来自主表 AMI 数据) ---
            \textbf{Ours (Full)} & \textbf{49.49} & - & 59.46 & 65.29 & 38.56 & 48.46 & 35.68   \\
            \midrule
            
            % --- Planner Ablation ---
            \multicolumn{8}{l}{\textit{Validation of Query Decomposer}} \\
            \hspace{2mm} w/o Query Planner & 46.39 & \textcolor{red}{-3.10} & 57.63 & 59.35 & 36.79 & 45.73 & 32.43 \\
            
            \multicolumn{8}{l}{\textit{Validation of Query Planner}} \\
            \hspace{2mm} w/o Query Planner & 44.68 & \textcolor{red}{-4.80} & 56.12 & 56.35 & 36.41 & 43.18 & 31.37 \\
            
            % --- Reflection Ablation ---
            \multicolumn{8}{l}{\textit{Validation of Reflection}} \\
            \hspace{2mm} w/o Reflection & 39.64 & \textcolor{red}{-9.84} & 51.32 & 49.52 & 31.10 & 39.83 & 26.45 \\

            % --- Tool Ablation ---
            \multicolumn{8}{l}{\textit{Validation of Action Space}} \\
            \hspace{2mm} w/o Tool: Filtering & 43.69 & \textcolor{red}{-5.79}           & 57.81 & 62.27 & 35.36 & 41.45 & 28.58  \\
            \hspace{2mm} w/o Tool: Semantic Search & 16.28 & \textcolor{red}{-33.21}    & 28.96 & 16.38 & 12.21 & 13.69 & 10.15   \\
            \hspace{2mm} w/o Tool: Graph Traversal & 38.21 & \textcolor{red}{-11.27}    & 45.57 & 38.62 & 32.31 & 41.25 & 33.31  \\
            \hspace{2mm} w/o Tool: Audio Access & 34.71 & \textcolor{red}{-14.78}       & 47.52 & 49.12 & 33.13 & 31.71 & 12.06  \\
            
            \bottomrule
        \end{tabular}%
    }
    \caption{\textbf{Ablation study on the AMI dataset.} We report the accuracy (\%) drop when specific components are removed. \textbf{$\Delta$}: Performance degradation compared to the full framework. }
    \label{tab:ablation}
\end{table}
\textbf{Ablation Study}
To validate the necessity of each component in our GRGA, we conduct an ablation study on the AMI dataset as a typical example in Table~\ref{tab:ablation}. From this table, we can see that removing any module results in varying degrees of performance degradation for our GRGA. Especially, the results of \textit{w/o Semantic Search} demonstrate that our approach can indeed solve the semantic understanding problem in long-form speech meeting scenario. The removal of \textit{Graph Traversal} indicates the importance of our designed multi-dimensional graph. More analysis can be found in Appendix \ref{tab:ablation_detail}.

% \paragraph{Impact of Cognitive Modules.} 
% Removing the \textbf{Query Planner} leads to a significant drop of 4.80\%, particularly in inferential tasks (-8.94\%). This confirms that complex queries (e.g., multi-hop reasoning) cannot be solved by single-step retrieval; explicit planning is essential for decomposing intents. 
% Most notably, the removal of the \textbf{Reflection} mechanism causes a sharp decline of 9.84\%. Without the ``verify'' loop, the agent is prone to hallucination, accepting the first retrieved chunk even if it is irrelevant. This validates our hypothesis that a POMDP-style feedback loop is critical for robustness.

% \paragraph{Impact of Graph Tools.}
% The \textbf{Graph Traversal} tool proves indispensable ($\Delta=-11.27\%$). When disabled, the agent degrades to a flat-text searcher, failing to aggregate scattered information via speaker edges ($e_{spk}$) or temporal edges ($e_{temp}$). 
% Furthermore, removing \textbf{Audio Access} severely impacts acoustic-aware questions (Accuracy 35.68\% $\to$ 12.06\%), demonstrating that text transcripts alone are insufficient for capturing paralinguistic cues like emotion or speaker overlap. 
% Finally, \textbf{Semantic Search} serves as the foundational entry point; its removal collapses the system ($\Delta=-33.21\%$), as the agent loses the ability to locate initial evidence nodes.

\textbf{Step Analysis} can be found in Appendix~\ref{step_analysis}.
\textbf{Case Study} can be found in Appendix~\ref{case_study}.
\textbf{Noise Sensitivity Analysis} can be found in Appendix~\ref{sec:noise_analysis}.

\section{Related Work}

\textbf{Speech and Meeting Question Answering.}
Early research in speech QA primarily focused on extracting answer spans from \textbf{short, single-speaker} speech segments, such as Spoken SQuAD~\cite{li2018spokensquadstudymitigating} and HeySQuAD~\cite{wu2023heysquad}. They normally focus on answering ranking \cite{longQA}, instead of generative QA. While recent datasets like AMI~\cite{jain2024insectidentificationwildami} and AliMeeting~\cite{alimeeting} provide rich multi-speaker meeting resources, they are predominantly used for ASR and diarization tasks rather than complex reasoning. Existing QA models on these datasets often treat transcripts as flat text sequences, neglecting the intricate temporal and interpersonal dependencies inherent in meetings. In contrast, our work targets \textbf{long-form meeting comprehension}, requiring models to navigate graph-structured dialogues involving multiple speakers and temporal dynamics.

\textbf{Large Speech Language Models.}
Recent advancements in multi-modal LLMs have enabled direct processing of audio inputs. However, current Speech LLMs face significant limitations in context length. For instance, Kimi-Audio~\cite{kimiteam2025kimiaudiotechnicalreport} is optimized for short clips ($ <30 $s) and suffers from truncation issues with longer inputs. Even state-of-the-art models like AudioFlamingo3~\cite{goel2025audioflamingo3advancing} are typically constrained to context of approximately $ 10 $ minutes. This bottleneck renders them unsuitable for long-form meeting analysis in the absence of external retrieval mechanisms.

\textbf{Retrieval-Augmented Generation (RAG).}
RAG has emerged as a standard paradigm for grounding LLMs in external knowledge~\cite{lewis2020rag}. Traditional ``Retrieve-then-Generate'' approaches rely on dense vector similarity to retrieve relevant contexts in a single pass. However, these methods struggle with \textit{multi-hop reasoning}, where the evidence is fragmented or requires logical deduction steps not captured by semantic similarity alone~\cite{hoprag,graphmpa}. Recent works have explored recursive retrieval or chain-of-thought prompting to address these limitations. Our framework advances this paradigm by formalizing retrieval as an \textbf{iterative planning process}, specifically tailored to handle the noisy and unstructured nature of ASR transcripts.

\textbf{LLM Agents and Tool Using.}
The emergence of LLMs has catalyzed the development of autonomous agents capable of using tools to solve complex tasks, exemplified by frameworks like ReAct~\cite{yao2022react} and Toolformer~\cite{schick2023toolformer}. While these agents demonstrate proficiency in open-domain tasks (e.g., web browsing, math)~\cite{nips/agentic-rag}, their application to \textbf{structured audio understanding} remains underexplored. We bridge this gap by defining a specialized \textbf{Action Space} for meeting analysis (e.g., time\_range\_search, hybrid\_search) and incorporating a \textbf{Reflection mechanism} modeled as a POMDP~\cite{pompd}, enabling the agent to self-correct in partially observable audio environments.

\section{Conclusion}
We present \textbf{GRGA}, an agentic framework that addresses the acoustic missing and context forgetting issues in long-form speech understanding. 
By structuring audio into a multimodal heterogeneous graph and formulating QA as a POMDP, we enable an agent to explicitly plan, navigate, and reason over complex interactions. 
Extensive experiments on our proposed \textbf{LongAudioQA} benchmarks demonstrate that our GRGA significantly outperforms both end-to-end Speech-LLMs and RAG-based SOTAs.

\section*{Acknowledgements}
This work was supported by Jiangsu Province Frontier Program Project (BF2025036), and Hong Kong RGC grant GRF \#15611021. They are also with Jiangsu Key Lab of Language Computing, Suzhou.

\section*{Limitations}
Our work has three main limitations. 

(1) Error Propagation: Since graph construction relies on upstream ASR and diarization, severe acoustic noise or speaker overlap may introduce artifacts into the reasoning graph. 

(2) Inference Latency: The iterative agentic workflow (planning and reflection) incurs higher computational cost than single-turn RAG, limiting real-time deployment. 

(3) Domain Specificity: Our evaluation focuses on structured meetings; generalizing to unstructured domains like movies or vlogs remains future work.

\section*{Ethics Statement}

\paragraph{Dataset Sourcing and Compliance.}
Our dataset is constructed based on three existing public corpora: AliMeeting~\cite{alimeeting}, AMI Meeting Corpus~\cite{jain2024insectidentificationwildami}, and DailyTalk~\cite{lee2022dailytalk}. We strictly adhere to the original licensing terms of these datasets (e.g., Creative Commons BY-NC-SA 4.0 and Apache License 2.0). We have reviewed the source data to ensure that no Personally Identifiable Information (PII) beyond what was originally consented to by the participants is exposed. For the AMI corpus, we utilize the specific split intended for academic research involving simulated scenarios, minimizing privacy risks associated with real-world private meetings.

\paragraph{Human Annotation and Fair Compensation.}
We employ highly qualified graduate students as annotators. We ensured that all participants were compensated at a rate significantly above the local minimum wage (approximately \$5 per hour), respecting fair labor standards. We also implemented strict protocols to protect annotators from exposure to any potentially harmful or offensive content, although the source datasets are generally free of such material.

\paragraph{Mitigation of LLM Bias and Hallucination.}
We acknowledge that utilizing Large Language Models (LLMs) for data generation may introduce inherent biases or hallucinations. To mitigate this, our human-in-the-loop pipeline strictly enforces factual consistency checks. We explicitly instructed annotators to discard questions that reinforce stereotypes or rely on hallucinated events not present in the audio. The high inter-annotator agreement ($\kappa = 0.95$) suggests that our verification process effectively filters out low-quality or biased generations.

\paragraph{Potential Societal Impact.}
The technology proposed in this paper aims to enhance productivity and accessibility by structuring long-form audio. However, we recognize the potential dual-use risk regarding unauthorized surveillance or privacy intrusion in workplace settings. We strongly advocate that the deployment of such meeting analysis tools must be accompanied by transparent consent from all recorded parties and robust data encryption measures. This dataset is released solely for research purposes to advance the field of interpretable audio understanding.

\newpage
\bibliography{custom}

\newpage
\appendix
% \section{Details of our GRGA}
% \begin{figure*}
%     \centering
%     \includegraphics[width=1\linewidth]{figure/method.pdf}
%     \caption{The overall architecture of our graph-based retrieval-generation agent \textbf{GRGA}. 
% The framework models long-form audio as a \textbf{Multi-dimensional Graph} (bottom-left) to capture semantic, temporal, and speaker dependencies. 
% Our GRGA Planning Process consists of: 
% (1) \textbf{Query Decomposition}: The user query is parsed into structured constraints (e.g., entity $e$, time $t$). 
% (2) \textbf{Planning}: The \textbf{Execution Planner} formulates a logical tool-use plan (e.g., invoking \texttt{time\_range\_search}). 
% (3) \textbf{Execution}: The \textbf{Tool Executor} interacts with the graph to retrieve evidence segments. 
% (4) \textbf{Synthesis}: The \textbf{Answer Synthesizer} produces a citation-backed response. 
% (5) \textbf{Reflection}: The \textbf{Self-Reflector} evaluates the answer's reliability. Crucially, a low score ($ S < \tau $) triggers a \textbf{feedback loop} (red arrow), updating the belief state for re-planning.}
%     \label{fig:method_details}
% \end{figure*}

\section{More Analysis in Ablation Study}

\begin{table}[h]
    \centering
    \renewcommand{\arraystretch}{1.2}
    \setlength{\tabcolsep}{3.5pt}
    
    \resizebox{\columnwidth}{!}{%
        \begin{tabular}{l | c c | c c c c c}
            \toprule
            \multirow{2}{*}{\textbf{Ablation Settings}} & \multicolumn{2}{c|}{\textbf{Overall}} & \multicolumn{5}{c}{\textbf{AMI }} \\
            \cmidrule(lr){2-3} \cmidrule(lr){4-8}
             & \textbf{Avg.} & \textbf{$\Delta$} & \textbf{Fact.} & \textbf{Infer.} & \textbf{Temp.} & \textbf{Summ.} & \textbf{Acou.} \\
            \midrule
            
            % --- Full Model (来自主表 AMI 数据) ---
            \textbf{Ours (Full)} & \textbf{49.49} & - & 59.46 & 65.29 & 38.56 & 48.46 & 35.68   \\
            \midrule
            
            % --- Planner Ablation ---
            \multicolumn{8}{l}{\textit{Validation of Query Decomposer}} \\
            \hspace{2mm} w/o Query Planner & 46.39 & \textcolor{red}{-3.10} & 57.63 & 59.35 & 36.79 & 45.73 & 32.43 \\
            
            \multicolumn{8}{l}{\textit{Validation of Query Planner}} \\
            \hspace{2mm} w/o Query Planner & 44.68 & \textcolor{red}{-4.80} & 56.12 & 56.35 & 36.41 & 43.18 & 31.37 \\
            
            % --- Reflection Ablation ---
            \multicolumn{8}{l}{\textit{Validation of Reflection}} \\
            \hspace{2mm} w/o Reflection & 39.64 & \textcolor{red}{-9.84} & 51.32 & 49.52 & 31.10 & 39.83 & 26.45 \\

            % --- Tool Ablation ---
            \multicolumn{8}{l}{\textit{Validation of Action Space}} \\
            \hspace{2mm} w/o Tool: Filtering & 43.69 & \textcolor{red}{-5.79}           & 57.81 & 62.27 & 35.36 & 41.45 & 28.58  \\
            \hspace{2mm} w/o Tool: Semantic Search & 16.28 & \textcolor{red}{-33.21}    & 28.96 & 16.38 & 12.21 & 13.69 & 10.15   \\
            \hspace{2mm} w/o Tool: Graph Traversal & 38.21 & \textcolor{red}{-11.27}    & 45.57 & 38.62 & 32.31 & 41.25 & 33.31  \\
            \hspace{2mm} w/o Tool: Audio Access & 34.71 & \textcolor{red}{-14.78}       & 47.52 & 49.12 & 33.13 & 31.71 & 12.06  \\
            
            \bottomrule
        \end{tabular}%
    }
    \caption{\textbf{Ablation study on the AMI dataset.} We report the accuracy (\%) drop when specific components are removed. \textbf{$\Delta$}: Performance degradation compared to the full framework. }
    \label{tab:ablation_detail}
\end{table}
To validate the necessity of each component in AudioGraph, we conduct an ablation study on the AMI dataset (Table~\ref{tab:ablation_detail}). 

\paragraph{Impact of Cognitive Modules.} 
Removing the \textbf{Query Planner} leads to a significant drop of 4.80\%, particularly in inferential tasks (-8.94\%). This confirms that complex queries (e.g., multi-hop reasoning) cannot be solved by single-step retrieval; explicit planning is essential for decomposing intents. 
Most notably, the removal of the \textbf{Reflection} mechanism causes a sharp decline of 9.84\%. Without the ``verify'' loop, the agent is prone to hallucination, accepting the first retrieved chunk even if it is irrelevant. This validates our hypothesis that a POMDP-style feedback loop is critical for robustness.

\paragraph{Impact of Graph Tools.}
The \textbf{Graph Traversal} tool proves indispensable ($\Delta=-11.27\%$). When disabled, the agent degrades to a flat-text searcher, failing to aggregate scattered information via speaker edges ($e_{spk}$) or temporal edges ($e_{temp}$). 
Furthermore, removing \textbf{Audio Access} severely impacts acoustic-aware questions (Accuracy 35.68\% $\to$ 12.06\%), demonstrating that text transcripts alone are insufficient for capturing paralinguistic cues like emotion or speaker overlap. 
Finally, \textbf{Semantic Search} serves as the foundational entry point; its removal collapses the system ($\Delta=-33.21\%$), as the agent loses the ability to locate initial evidence nodes.

\section{Evidence Precision Recall Analysis}

\begin{table}[h!]
    \centering
    \renewcommand{\arraystretch}{1.15}
    \setlength{\tabcolsep}{3.5pt} % 极限压缩列宽
    
    \resizebox{\columnwidth}{!}{%
        \begin{tabular}{l cc cc cc}
            \toprule
            \multirow{2}{*}{\textbf{Method}} & \multicolumn{2}{c}{\textbf{DailyTalk}} & \multicolumn{2}{c}{\textbf{AMI}} & \multicolumn{2}{c}{\textbf{AliMeeting}} \\
            \cmidrule(lr){2-3} \cmidrule(lr){4-5} \cmidrule(lr){6-7}
             & P & R & P & R & P & R \\
            \midrule
            Qwen3-Omni      & 32.6 & 24.6 & 25.7 & 27.3 & 29.3 & 28.2 \\
            \quad + Text RAG     & 67.3 & 69.1 & 56.9 & 65.2 & 72.4 & 68.3 \\
            \quad + Audio RAG     & 61.6 & 63.6 & 45.2 & 42.7 & 57.3 & 49.6 \\
            \textbf{\quad + Ours} & \textbf{84.2} & \textbf{77.2} & \textbf{76.2} & \textbf{69.7} & \textbf{79.1} & \textbf{74.1} \\
            \bottomrule
        \end{tabular}%
    }
    \caption{\textbf{Citation accuracy comparison.} Results are reported in \% with a $\pm 2$s tolerance. P: Precision, R: Recall.}
    \label{tab:pr_tab}
\end{table}

Table \ref{tab:pr_tab} compares the evidence citation accuracy across three datasets, demonstrating that our method consistently outperforms both the vanilla model and RAG baselines.

\section{Step Analysis}
\label{step_analysis}
\begin{figure}[h!]
    \centering
    \includegraphics[width=1\linewidth]{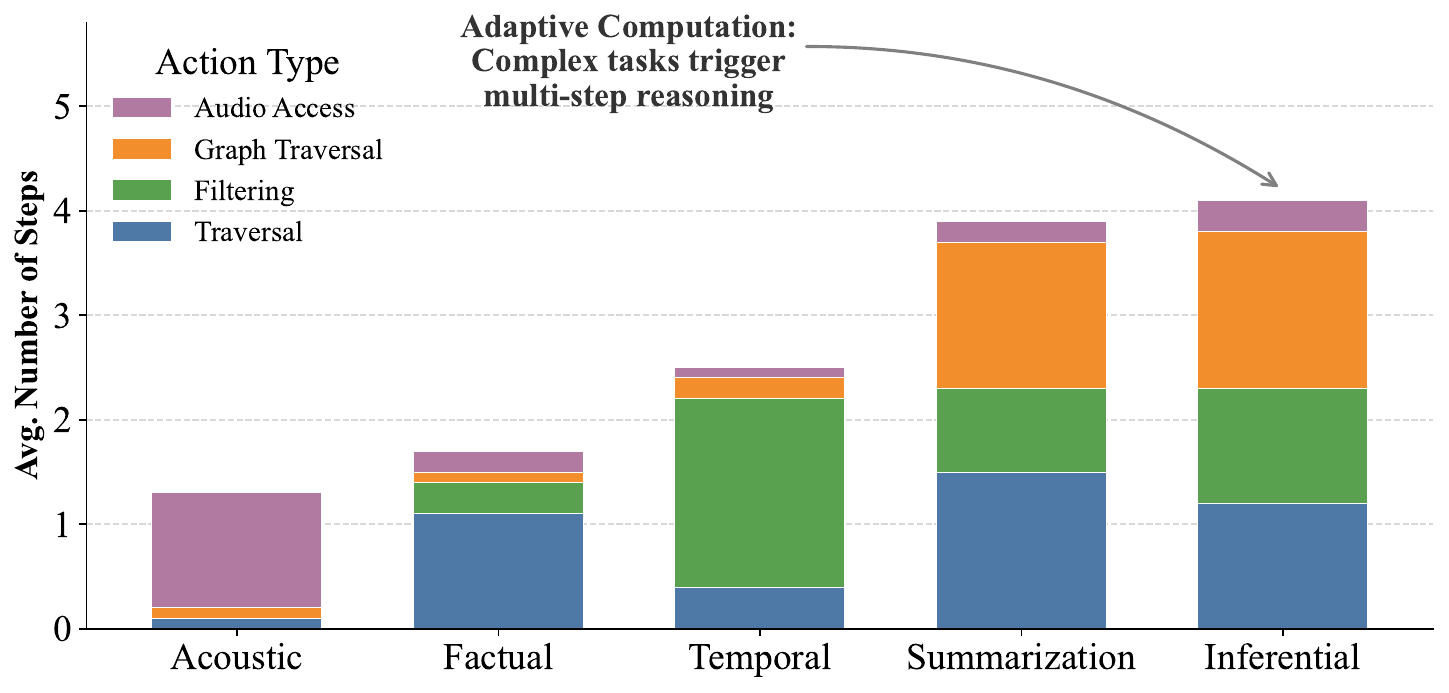}
    \caption{\textbf{Average reasoning steps across different question types.} 
The agent exhibits \textit{adaptive computation}: solving simple Factual queries requires minimal steps, while complex Inferential and Temporal queries trigger deeper reasoning chains. 
The stacked colors illustrate the distribution of tool usage, showing heavy reliance on \categorylabel{c_gt}{Graph Traversal} for multi-hop reasoning.}
    \label{fig:step_analysis}
\end{figure}
\paragraph{Complexity-Aware Reasoning.}
The results demonstrate a clear correlation between question difficulty and planning depth. 
For \textbf{Factual} queries (e.g., \textit{``What is the budget?''}), the agent adopts a ``shortcut'' strategy, typically resolving the intent in just 1.8 steps using primarily \texttt{Semantic Search}. This indicates high efficiency.
In contrast, \textbf{Inferential} and \textbf{Temporal} queries trigger significantly longer trajectories (avg. $>$4 steps). This confirms that the agent is actively performing multi-hop reasoning—iteratively traversing $e_{temp}$ or $e_{spk}$ edges to aggregate scattered evidence, rather than relying on a single retrieval pass.

\paragraph{Tool Usage Distribution.}
The stacked breakdown reveals task-specific tool preferences. 
\textbf{Summarization} tasks show a dominant usage of \texttt{Filter} tools (Green) to isolate specific time ranges or speakers.
Crucially, \textbf{Acoustic-aware} questions exhibit a high frequency of \texttt{Audio Access} (Red) calls in the final steps, validating that the agent correctly learns to ``listen'' to the raw audio only when textual transcripts are insufficient (e.g., verifying emotion).

\begin{figure*}[t]
    \centering
    \includegraphics[width=1.0\linewidth]{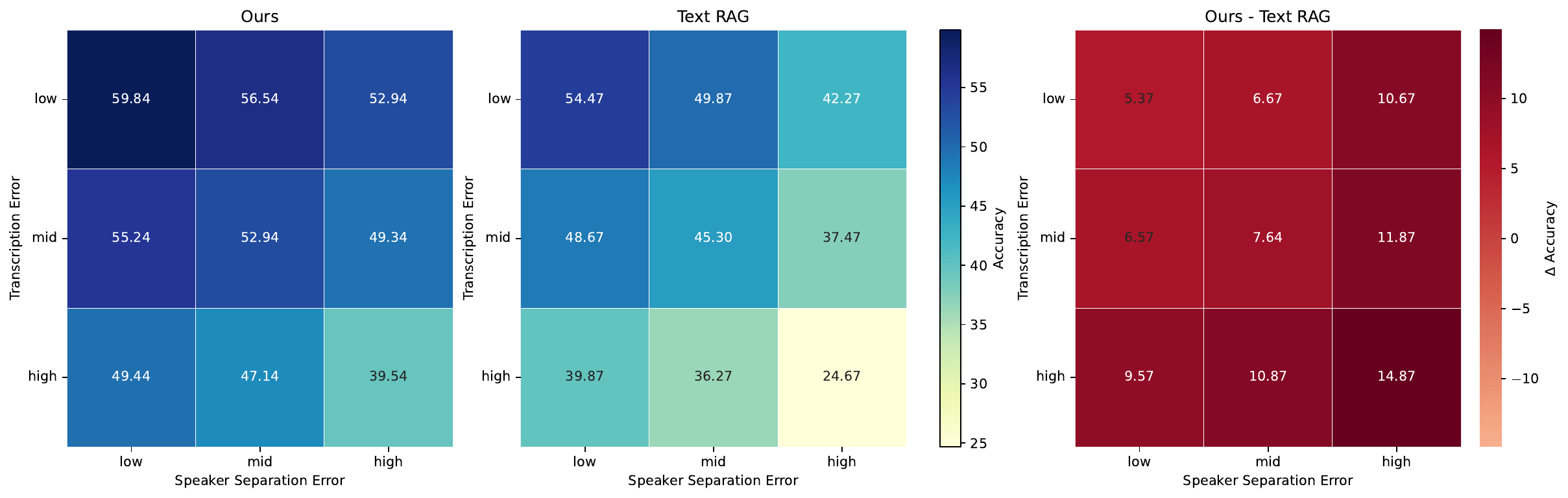} 
    \caption{\textbf{Noise Sensitivity Analysis.} QA Accuracy comparison under varying upstream error rates. The rightmost panel highlights the widening performance gap ($\Delta$ Accuracy), showing that GRGA's advantage grows in noisier environments.}
    \label{fig:sensitivity_heatmap}
\end{figure*}
\section{Human Evaluation.}

\begin{table}[t]
\centering
\resizebox{\linewidth}{!}{
\begin{tabular}{ll|ccc|c|c}
\toprule
\multirow{2}{*}{\textbf{Dataset}} & \multirow{2}{*}{\textbf{Method}} & \multicolumn{3}{c|}{\textbf{Human Ratings (1-5 Scale)}} & \textbf{Avg.} & \textbf{IAA} \\
\cmidrule(lr){3-5}
 & & \textbf{Corr.} & \textbf{Grd.} & \textbf{Coh.} & & \\
\midrule
\multirow{2}{*}{\textbf{AMI}} 
 & Qwen3-Omni & 1.79 & 2.12 & 3.76 & 2.56 & \multirow{2}{*}{0.78} \\
 & \textbf{Ours} & \textbf{2.25} & \textbf{3.25} & \textbf{4.35} & \textbf{3.28} & \\
\midrule
\multirow{2}{*}{\textbf{AllMeeting}} 
 & Qwen3-Omni & 1.93 & 2.24 & 4.32 & 2.83 & \multirow{2}{*}{0.75} \\
 & \textbf{Ours} & \textbf{2.36} & \textbf{3.51} & \textbf{4.37} & \textbf{3.41} & \\
\midrule
\multirow{2}{*}{\textbf{DailyTalk}} 
 & Qwen3-Omni & 3.43 & 3.72 & 4.12 & 3.76 & \multirow{2}{*}{0.82} \\
 & \textbf{Ours} & \textbf{4.11} & \textbf{4.23} & \textbf{4.37} & \textbf{4.24} & \\
\midrule
\multirow{3}{*}{\textbf{Overall}} 
 & Average Qwen3-Omni & 2.38 & 2.69 & 4.07  & \multirow{3}{*}{-} & \multirow{3}{*}{-} \\
 & \textbf{Average Ours} & \textbf{2.91} & \textbf{3.66} & \textbf{4.36} &   & \\
 & \textbf{Improvement (\%)} & \textbf{+0.52} & \textbf{+0.97} & \textbf{+0.30} &   & \\
\bottomrule
\end{tabular}
}
\caption{Human evaluation results across different datasets. Models are assessed on three dimensions: Correctness (Corr.), Groundedness (Grd.), and Coherence (Coh.) using a 1-5 scale. We also report average scores (Avg.) and Inter-Annotator Agreement (IAA).}
\label{tab:human_eval}
\end{table}

To validate real-world performance, we conducted a blind evaluation on 150 queries randomly sampled across five question types from AMI, AliMeeting, and DailyTalk. Three graduate student annotators assessed the responses (IAA=0.78 on avg.). Results in Table~\ref{tab:human_eval} shows:

\paragraph{The ``Fluency vs. Factualit'' Gap.} 
While the baseline (Qwen3-Omni) maintains high \textit{Coherence} (4.07), its low \textit{Groundedness} score (2.69) indicates a tendency to generate fluent but hallucinated content. In contrast, our method achieves a massive \textbf{+0.97 improvement in Groundedness}. This confirms that our approach does not merely generate text but accurately anchors answers to precise timestamps and speakers, effectively mitigating hallucinations.

\paragraph{Grounding Drives Correctness.}
There is a clear positive correlation between groundedness and correctness. By explicitly locating evidence, our method achieves a \textbf{+0.52 gain in Correctness}. This demonstrates that superior localization capabilities directly translate to more factually accurate answers, particularly for reasoning-heavy questions.

\paragraph{Robustness in Complex Scenarios.}
The performance gap is most pronounced on challenging datasets like AMI and AliMeeting (noisy, multi-party interactions) compared to the cleaner DailyTalk. Our method maintains distinct advantages in these ``hard'' settings (e.g., \textbf{+1.13 Groundedness on AMI}), proving its robustness where standard omni-models struggle with speaker attribution and temporal reasoning.

\label{case_study}
\begin{table*}[t]
\small
\centering
\renewcommand{\arraystretch}{1.25}
\resizebox{\textwidth}{!}{
\begin{tabular}{l p{0.25\linewidth} p{0.65\linewidth}}
\toprule
\textbf{Stage} & \textbf{Module / Action} & \textbf{Execution Details \& Internal State} \\
\midrule
\multicolumn{3}{l}{\cellcolor{gray!10}\textbf{User Query:} \textit{``What is the team's general attitude to the project deadline and workload?''}} \\

% --- PHASE 1: INITIAL ATTEMPT ---
\midrule
\multirow{6}{*}{\textbf{Phase 1}} & \textbf{Intent Analysis} & \textbf{Entities:} [`team', `project deadline', `workload'] \quad \textbf{Concept:} [`general attitude'] \\
\cmidrule(lr){2-3}
& \textbf{Plan Execution} & 
\textbf{1. Hybrid Search:} query=``team project deadline workload'' ($\rightarrow$ 10 hits) \newline
\textbf{2. Traverse Relations:} Expand context ($\rightarrow$ 64 nodes found). \\
\cmidrule(lr){2-3}
& \textbf{Draft Answer} & \textit{``The team generally views the deadline as tight... members feeling urgency.''} \\
& & \textcolor{gray}{\small (Generic summary, lacks specific details)} \\
\cmidrule(lr){2-3}
& \cellcolor{bg_fail}\textbf{Reflection (Critic)} & 
\cellcolor{bg_fail}\textbf{Assessment:} \textcolor{red}{\xmark \textbf{Confidence: 2/5}} \newline
\cellcolor{bg_fail}\textit{Reasoning:} ``Supported is False. Need reverse confirmation for `tight' and `work-intensive' claims.'' \newline
\cellcolor{bg_fail}$\hookrightarrow$ \textbf{Decision:} Trigger \textsc{Re-Plan}. \\

% --- PHASE 2: RE-PLANNING ---
\midrule
\multirow{8}{*}{\textbf{Phase 2}} & \textbf{Re-Plan Strategy} & \textbf{New Goal:} Validate sentiment strength (e.g., stress, specific complaints). \newline
\textbf{Tool:} \texttt{hybrid\_search(query=``project deadline focused effort'', alpha=0.6)} \\
\cmidrule(lr){2-3}
& \textbf{Evidence Retrieval} & 
The new search strategy retrieves highly specific emotional utterances: \newline
$\bullet$ \texttt{utt180}: ``...only got \textbf{six days} to do the topic segmentation'' \newline
$\bullet$ \texttt{utt225}: ``shall we \textbf{change the deadlines}?'' \newline
$\bullet$ \texttt{utt349}: ``that's a \textbf{hassle} too it's a \textbf{lot of work}'' \newline
$\bullet$ \texttt{utt553}: ``i wanna do it \textbf{superficially}'' \\
\cmidrule(lr){2-3}
& \cellcolor{bg_success}\textbf{Final Response} & 
\cellcolor{bg_success}\textit{``The team's attitude indicates the project deadline is perceived as tight... One member states they \textbf{only have six days} [\texttt{utt180}]. Another asks \textbf{`shall we change the deadlines?'} [\texttt{utt225}]. Another calls it \textbf{`a hassle'} [\texttt{utt349}]... expresses a preference to handle it \textbf{`superficially'} [\texttt{utt553}].''} \\
\cmidrule(lr){2-3}
& \textbf{Final Reflection} & \textbf{Assessment:} \textcolor{green!60!black}{\cmark \textbf{Confidence: 5/5}} \quad \textbf{Status:} Supported. \\

\bottomrule
\end{tabular}
}
\caption{\textbf{Case Study on Self-Correction.} Initially, the model retrieved broad context but generated a low-confidence (2/5), generic answer. The \textit{Reflection} module detected the lack of specific evidence and triggered a re-planning step focusing on ``focused effort.'' This allowed the model to retrieve concrete complaints (e.g., ``six days,'' ``hassle,'' ``superficially''), resulting in a highly grounded (5/5) final response.}
\label{tab:reflection_case_study}
\end{table*}

\section{Noise Sensitivity Analysis}
\label{sec:noise_analysis}

To evaluate the robustness of our proposed GRGA against upstream errors, we conducted a sensitivity analysis by simulating varying levels of Word Error Rate (WER) and Diarization Error Rate (DER). 
We compare our model with the baseline Text RAG system across a $3 \times 3$ grid of noise conditions.

We define three noise levels for both ASR and Diarization components:

\textbf{Low (Oracle):} We use Ground Truth (GT) transcripts and speaker labels to simulate an ideal scenario.

\textbf{Mid (Standard):} We utilize the outputs from our standard pipeline (as described in Section \ref{sec:implementation}), representing a realistic deployment scenario (DER $\approx$ 17.6\%, WER $\approx$ 18.8\%).

\textbf{High (Simulated Noise):} 
For \textit{High WER}, we utilize a lower-performance lightweight ASR model (openai/whisper-small ~\cite{whisper}) to generate transcripts with high error rates (WER $\approx$ 33.6\%).
For \textit{High DER}, we simulated noise by randomly shuffling a subset of speaker labels within a meeting while preserving timestamp boundaries. This represents a worst-case scenario where speaker identity information is highly unreliable (DER $\approx$ 52.4\%). 

Figure \ref{fig:sensitivity_heatmap} visualizes the impact of noise, revealing distinct behaviors:

\textbf{Graceful Degradation.}
Text RAG suffers catastrophic drops as noise increases, falling from $54.47\%$ to $24.67\%$ in the \textit{High/High} setting. This underscores the fragility of standard RAG when explicit text or speaker cues are noisy. Conversely, GRGA demonstrates graceful degradation, retaining $39.54\%$ accuracy even under the harshest conditions.

\textbf{The ``Noise Buffer'' Effect.}
Notably, the performance gap between GRGA and the baseline widens from $+5.37\%$ (Low/Low) to $+14.87\%$ (High/High). This indicates that our graph structure acts as an effective noise buffer. By leveraging topological connectivity and temporal constraints, the agent can compensate for corrupted signals, reducing the system's reliance on perfect transcription and diarization.

\section{Case Study}
Table \ref{tab:reflection_case_study} presents a case study illustrating our method's self-correction capability, where the agent initially generates a low-confidence response but, triggered by the reflection module, re-plans to retrieve specific evidence (e.g., concrete complaints about deadlines), ultimately producing a highly grounded and accurate answer.

\section{Experimental Setting Details}
\label{experimental_details}
\subsection{Baselines}

\textbf{Qwen3-Omni.}
Qwen3-Omni~\cite{Qwen3-Omni} is a state-of-the-art (SOTA) end-to-end omni-modal model that processes interleaved inputs (e.g., text and audio) and generates text responses. It supports speech inputs of up to 40 minutes. The model adopts a Mixture-of-Experts (MoE) architecture with 30B parameters, of which 3B are activated per inference. Notably, it achieves performance comparable to top-tier proprietary models (e.g., Gemini 2.5 Pro~\cite{comanici2025gemini25pushingfrontier}) across a wide range of audio understanding benchmarks.

\textbf{AudioFlamingo3.} 
We select \textbf{Audio Flamingo 3}~\cite{goel2025audioflamingo3advancing} as a SOTA audio-language model. 
Built upon Whisper encoder and 7B LLM, it employs an ``on-demand thinking'' mechanism to facilitate chain-of-thought reasoning. 
Crucially, its architecture processes inputs via 30-second windows with a strict context cap of 10 minutes. 
This baseline serves to quantify the performance degradation of context-constrained models when applied to long meeting scenarios.

\textbf{MiMo-Audio.}
MiMo-Audio ~\cite{coreteam2025mimoaudio} is a SOTA end-to-end audio language model. It combines a 1.2B audio tokenizer with a 7B LLM. The model supports a 128k context window and downsamples audio tokens to 6.25\,Hz, enabling efficient processing of long sequences. Crucially, its instruction-tuning stage incorporates ``thinking'' mechanisms, allowing chain-of-thought reasoning directly over audio inputs. Comparing against MiMo-Audio helps assess whether our graph-based structural reasoning provides advantages beyond large-scale end-to-end pretraining.

\textbf{Text RAG Baseline.}
We implement a standard dense retrieval baseline using \textbf{BGE-M3}~\cite{chen2024bgem3embeddingmultilingualmultifunctionality} embeddings. 
Meeting transcripts are segmented into utterances with prepended metadata (e.g., \texttt{[time] SPK: text}). 
Given a query $q$, we retrieve the top-$k$ ($k=10, \lambda=0.25$) most relevant segments based on cosine similarity, get text result and corresponding audio clips. 
Crucially, to maintain discourse coherence, retrieved segments are temporally reordered before being fed into the \textbf{Speech-LLM}. 
This baseline represents the conventional approach to handling long meetings without graph-based structural reasoning.

\textbf{Audio RAG Baseline.}
We construct a cross-modal retrieval baseline using \textbf{CLASP}~\cite{CLASP}, which aligns audio and text in a shared semantic space. 
The long audio is segmented into clips based on speaker diarization timestamps. 
Given a text query $q$, we retrieve the top-$k$ most relevant audio clips via cosine similarity between the query embedding and CLASP-encoded audio embeddings. 
These retrieved raw audio clips, along with their speaker metadata, are then fed directly into the \textbf{Speech-LLM} (same backbone as ours) to generate the response. 
This baseline evaluates the efficacy of retrieving raw acoustic features versus our proposed graph-based semantic navigation.

\subsection{Evaluation Metric: Semantic Accuracy}

To rigorously assess the reasoning capabilities of our model, we report \textbf{Semantic Accuracy} across all datasets. 
Traditional n-gram metrics (e.g., BLEU, Exact Match) often fail to capture the true validity of generated responses, as they penalize correct answers that differ in lexical surface forms from the ground truth. 
Drawing upon recent methodologies in complex reasoning evaluation \cite{mishra-etal-2025-investigating, shui-etal-2023-comprehensive}, we move beyond rigid string matching and establish a robust, automated evaluation pipeline that prioritizes semantic equivalence and logical soundness.

Specifically, we employ a Strong LLM (LLM-as-a-Judge) to approximate human-level judgment. 
Formally, let $\mathcal{D} = \{(q_i, a_i)\}_{i=1}^{N}$ denote the evaluation dataset, where $q_i$ is the query and $a_i$ is the ground-truth answer. Let $\hat{a}_i$ represent the model-generated response. 
We define a semantic indicator function $\mathbb{I}(\cdot)$ parameterized by an external judge $\mathcal{J}$ (e.g., GPT-OSS-120B \cite{openai2025gptoss120bgptoss20bmodel}):
\begin{equation}
    s_i = \mathcal{J}(q_i, a_i, \hat{a}_i) \in \{0, 1\},
\end{equation}
where $s_i = 1$ if and only if $\mathcal{J}$ determines that $\hat{a}_i$ entails the same semantic information as $a_i$, and $0$ otherwise. 
The judge is prompted to ignore stylistic differences and focus strictly on factual consistency and reasoning correctness. 
Finally, the Semantic Accuracy is computed as the expectation of correct judgments:
\begin{equation}
    \text{Accuracy} = \frac{1}{N} \sum_{i=1}^{N} s_i \times 100\%.
\end{equation}
This approach ensures a fair comparison by validating whether the model successfully retrieves and reasons over the core knowledge, regardless of its output phrasing.

\section{Implementation Details}
\label{sec:implementation}

All the methods that did not explicitly state the model used Qwen3-Omni~\cite{Qwen3-Omni}.

\textbf{Graph Construction Setup.} For the initial audio processing, we utilize the \textbf{FireRedASR-AED} ~\cite{FireRedASR} model for Automatic Speech Recognition (ASR) to transcribe the raw audio meeting data. To obtain precise word-level timestamps (i.e., $ t_{start} $ and $ t_{end} $), we employ the \textbf{Montreal Forced Aligner (MFA)}.

\textbf{Speaker Verification:} We use a pre-trained \textbf{ERes2NetV2} ~\cite{ERes2NetV2} model to extract speaker embeddings. The cosine similarity threshold for speaker clustering is set to $ \eta = 0.8 $ based on validation set performance.
\textbf{Edge Construction:} The temporal window for establishing ``Reply-To'' edges is set to $ \Delta t = 5 $ seconds.
\textbf{Attribute Extraction:} We employ Qwen3-Omni (via vllm API) to extract speaker profiles and resolve coreference chains during the node enrichment phase.

\textbf{Agent Planning Configuration}
The core planning mechanism of GRGA depends on specific hyper-parameters governing the exploration-exploitation trade-off:
\textbf{Backbone Model:} The Policy Network $ \pi_\theta $ is parameterized by \textbf{Qwen3-Omni}, operated with a temperature of $ 0 $ to ensure deterministic reasoning paths.
\textbf{Reward Function:} As defined in Eq. (9), the self-reflection verification threshold is set to $ \tau = 4 $ (1-5 Scale). The penalty factor for failed verification is set to $ \beta = 0.5 $.
\textbf{Search Constraints:} The maximum depth for the reasoning tree is limited to $ K_{max} = 5 $ steps. If the agent fails to reach a verified conclusion within these steps, the process terminates and returns the current best candidate.

All experiments were conducted on a server cluster equipped with 8 $\times$ Ascend 910B (64GB) GPUs. 

\subsection{Text RAG Baseline.}

To benchmark our method against a standard text-based retrieval pipeline, we implement a dense Retrieval-Augmented Generation (RAG) baseline over meeting transcripts.

\textit{Context construction.} We segment the meeting transcript into individual utterances and treat each utterance as one retrieval unit. Each unit is formatted with temporal and speaker metadata:
\[
u_i=\texttt{[start}_i\texttt{ - end}_i \texttt{] SPEAKER\_ID: text}_i.
\]

\textit{Dense retrieval.} We use BGE-M3~\cite{chen2024bgem3embeddingmultilingualmultifunctionality} as the embedding backbone. Given a query $q$, we compute embeddings for the query and each utterance:
\[
\mathbf{e}_q=f_{\text{BGE}}(q) \in \mathbb{R}^{1024},\qquad \mathbf{e}_i=f_{\text{BGE}}(u_i) \in \mathbb{R}^{1024}.
\]
We apply L2 normalization to enable cosine-based scoring:
\[
\hat{\mathbf{e}}=\frac{\mathbf{e}}{\lVert \mathbf{e}\rVert_2},\qquad
s_i=\cos(\hat{\mathbf{e}}_q,\hat{\mathbf{e}}_i)=\hat{\mathbf{e}}_q^\top \hat{\mathbf{e}}_i.
\]
We rank all utterances by $s_i$ and retrieve the top-$k$ segments ($k=10$) that satisfy a minimum similarity threshold $\lambda=0.25$:
\[
\mathcal{C}_{\text{sim}}=\operatorname{TopK}\Big(\{u_i \mid s_i\ge\lambda\},\,k\Big).
\]

\textit{Temporal reordering.} To preserve discourse coherence, we reorder the retrieved utterances by their start time before forming the final context:
\[
\mathcal{C}=\operatorname{SortByTime}(\mathcal{C}_{\text{sim}}).
\]

\textit{Generation.} We concatenate the reordered segments as context $c=\operatorname{Concat}(\mathcal{C})$ and prompt the generator to answer \emph{strictly based on the retrieved context}. We use the same backbone LLM as in our main method for a fair comparison:
\[
y=\text{LLM}(q,c).
\]

\subsection{Audio RAG Baseline.}
Given a text query $q$ and a long audio recording $A$, we build a retrieval-augmented generation (RAG) baseline based on CLASP~\cite{CLASP}, a multilingual audio-text representation model that maps raw speech into a shared 768-dimensional semantic space.

\textbf{Audio segmenttation.}

We segment the long audio $A$ into $K$ shorter clips $\{a_k\}_{k=1}^{K}$, by speaker diarization.

\textbf{Embedding computation.}
We compute the query embedding and the audio clip embeddings using CLASP:
\begin{align}
\mathbf{e}_q &= f_{\text{text}}(q) \in \mathbb{R}^{768}, \\
\mathbf{e}_k &= f_{\text{audio}}(a_k) \in \mathbb{R}^{768}, \quad k=1,\dots,K,
\end{align}
where $f_{\text{audio}}(\cdot)$ encodes speech (e.g., via HuBERT and spectrogram encoders with a fusion module), and $f_{\text{text}}(\cdot)$ encodes text into the same representation space.

\textbf{Retrieval.}
We rank audio clips by cosine similarity and select the most relevant clip (or top-$M$ clips):
\begin{align}
s_k &= \cos(\mathbf{e}_q, \mathbf{e}_k) = \frac{\mathbf{e}_q^\top \mathbf{e}_k}{\|\mathbf{e}_q\|_2\,\|\mathbf{e}_k\|_2}, \\
k^\star &= \arg\max_{k \in \{1,\dots,K\}} s_k.
\end{align}
Optionally, we retrieve $\mathcal{K}=\text{TopM}(\{s_k\}_{k=1}^K)$ for multi-evidence prompting.

\textbf{Generation.}
We feed the query and retrieved evidence to a Speech LLM to generate the final response:
\begin{align}
y &= \text{LLM}\big(q, \{a_k\}_{k \in \mathcal{K}}\big),
\end{align}
where the evidence is provided, contains retrieval speaker diarization and corresponding audio clips.

\section{Multi-dimensional Meeting Database Construction Algorithm}

\paragraph{Complexity.}
Let $M$ be the number of VAD clips, and let $F_m$ be the number of acoustic frames in clip $m$.
Let $N_w$ be the total number of word tokens after forced alignment, $N_s$ the total number of diarization segments, and $N_u$ the number of utterances.

\textbf{ASR inference} and \textbf{CTC forced alignment} are linear in the input length, costing
$\mathcal{O}\!\left(\sum_{m=1}^{M} F_m\right)$ time.
\textbf{Speaker diarization} requires embedding extraction over frames plus clustering over segments; in practice this is
$\mathcal{O}\!\left(\sum_{m=1}^{M} F_m\right)$ for feature extraction, and an additional clustering cost by AHC is $\mathcal{O}(N_s^2)$.
\textbf{Speaker assignment} via overlap matching costs $\mathcal{O}(N_w+N_s)$ using the two-pointer implementation in Alg.~\ref{alg:db_construction} (a naive all-pairs matching would be $\mathcal{O}(N_wN_s)$).
Building utterances from word streams is $\mathcal{O}(N_w)$, and adding temporal edges in the meeting graph is $\mathcal{O}(N_u)$.

The space complexity is dominated by storing word- and utterance-level annotations, i.e., $\mathcal{O}(N_w+N_u)$, plus the storage overhead of the text and vector indices for utterance retrieval.

\onecolumn
\begin{algorithm}[h]
\caption{Multi-dimensional Meeting Database Construction}
\label{alg:db_construction}

\KwIn{Long meeting audio $X$; session id $sid$; hyper-parameters
$\theta=\{\texttt{max\_clip\_len}, \texttt{vad\_pad}, \beta_{\text{sil}}, \alpha_{\text{ovlp}}, K\}$}
\KwOut{Structured database $\mathcal{D}$; meeting graph $G=(V,E)$; indices $\mathcal{I}$}

$X \leftarrow \textsc{Preprocess}(X)$ \tcp*[r]{resample/mono/normalize}
$\mathcal{C} \leftarrow \textsc{FSMN\_VAD}(X,\texttt{vad\_pad},\texttt{max\_clip\_len})$ \tcp*[r]{clips with absolute spans}

Initialize tables \textsc{Clips}, \textsc{Words}, \textsc{SpkSegs}, \textsc{Utterances}\;
Initialize graph $G=(V,E)$\;

\ForEach(\tcp*[f]{process each VAD clip}){clip $c \in \mathcal{C}$}{
    \textsc{Clips} $\leftarrow$ \textsc{Clips} $\cup \{(sid,c.id,c.t_s,c.t_e,c.wav)\}$\;

    $(T,\mathbf{P}) \leftarrow \textsc{ASR}(c.wav)$ \tcp*[r]{transcript and CTC posteriors}
    $\mathcal{W} \leftarrow \textsc{CTC\_ForcedAlign}(T,\mathbf{P})$\;
    \tcp*[r]{$\mathcal{W}=\{(w_i,a_i,b_i,p_i)\}$ word-level timestamps}

    $\mathcal{S} \leftarrow \textsc{Diarize}(c.wav)$ \tcp*[r]{$\mathcal{S}=\{(spk_j,s_j,e_j,emb_j,conf_j)\}$}
    $\mathcal{S} \leftarrow \textsc{ClusterAndSmooth}(\mathcal{S})$ \tcp*[r]{merge/smooth speaker segments}
    \textsc{SpkSegs} $\leftarrow$ \textsc{SpkSegs} $\cup \{(sid,c.id,\mathcal{S})\}$\;

    $\widetilde{\mathcal{W}} \leftarrow \textsc{AssignSpeakerToWordsFast}(\mathcal{W},\mathcal{S},\alpha_{\text{ovlp}})$\;
    \textsc{Words} $\leftarrow$ \textsc{Words} $\cup \{(sid,c.id,\widetilde{\mathcal{W}})\}$\;

    $\mathcal{U} \leftarrow \textsc{BuildUtterances}(\widetilde{\mathcal{W}},\beta_{\text{sil}})$ \tcp*[r]{group into utterances}
    \ForEach{utterance $u \in \mathcal{U}$}{
        \textsc{Utterances} $\leftarrow$ \textsc{Utterances} $\cup \{(sid,u.id,u.spk,u.t_s,u.t_e,u.text,u.audio\_ref,u.conf)\}$\;
        $V \leftarrow V \cup \{u.id\}$ \tcp*[r]{utterance nodes in $G$}
    }
}

$E \leftarrow E \cup \textsc{AddTemporalEdges}(\textsc{Utterances})$ \tcp*[r]{time adjacency}
$E \leftarrow E \cup \textsc{AddSpeakerEdges}(\textsc{Utterances})$ \tcp*[r]{same-speaker links (optional)}
$E \leftarrow E \cup \textsc{AddEntityOrTopicEdges}(\textsc{Utterances})$ \tcp*[r]{optional NER/topic links}

$\mathcal{I}_{bm25} \leftarrow \textsc{BuildBM25Index}(\textsc{Utterances.text})$\;
$\mathcal{I}_{vec} \leftarrow \textsc{BuildVectorIndex}(\textsc{Utterances},K)$\;
$\mathcal{I} \leftarrow \{\mathcal{I}_{bm25},\mathcal{I}_{vec}\}$\;

$\mathcal{D} \leftarrow \{\textsc{Clips},\textsc{Words},\textsc{SpkSegs},\textsc{Utterances},\mathcal{I}\}$\;
\Return $\mathcal{D},G,\mathcal{I}$\;
\end{algorithm}
\section{Retrieval-Generation Agent Algorithm}
\begin{algorithm}[H]
\caption{GRGA with Formal Notation}
\label{alg:grga_formal}
\SetKwInput{Input}{Input}
\SetKwInput{Output}{Output}

\Input{$Q \in \mathcal{Q}$, $\mathcal{G} = (V, E, X)$}
\Output{$\hat{A} \in \mathcal{A}$ with $Cite \subseteq V \times T$}

\tcp{\textbf{Decomposition:} $f_{dec}: \mathcal{Q} \to \mathcal{C}$}
$\mathcal{C} = f_{dec}(Q) = \{c_e, c_c, c_t, c_m\}$\\
$b_0 = H_0 = \{Q, \mathcal{C}\}$; \quad $k \leftarrow 0$

\tcp{\textbf{POMDP Loop}}
\While{$k < K_{max}$}{
    \tcp{\textbf{Policy:} $\pi_\theta: \mathcal{B} \to \Delta(\mathcal{A}^*)$}
    $\mathcal{P}_k \sim \pi_\theta(\cdot \mid b_k)$ where $b_k = b_0 \cup H_k$
    
    \If{$\mathcal{P}_k = \bot$}{\Return $\bot$}
    
    \tcp{\textbf{Transition:} $T: \mathcal{S} \times \mathcal{A} \to \Delta(\mathcal{S})$}
    $o_k = \bigoplus_{op \in \mathcal{P}_k} \text{Exec}(op, \mathcal{G})$ where $o_k \in \mathcal{O}$
    
    \tcp{\textbf{Belief Update:} $\psi: \mathcal{B} \times \mathcal{A} \times \mathcal{O} \to \mathcal{B}$}
    $b_{k+1} = \psi(b_k, \mathcal{P}_k, o_k) = b_k \cup \{\mathcal{P}_k, o_k\}$
    
    \tcp{\textbf{Synthesis:} $f_{syn}: \mathcal{Q} \times \mathcal{B} \to \mathcal{A} \times \mathcal{C}$}
    $(\hat{A}_k, Cite_k) = f_{syn}(Q, b_{k+1})$
    
    \tcp{\textbf{Reward:} $R: \mathcal{S} \times \mathcal{A} \to \mathbb{R}$}
    $s_{ver} = f_{ref}(Q, \hat{A}_k, o_k) \in [0,1]$\\
    $r_k = \begin{cases} 
    1 & \text{if } s_{ver} \ge \tau \\
    -\beta & \text{otherwise}
    \end{cases}$
    
    \eIf{$r_k > 0$}{
        \Return $(\hat{A}_k, Cite_k)$
    }{
        $H_{k+1} = H_k \cup \{(\mathcal{P}_k, o_k, \hat{A}_k, r_k)\}$; \quad $k \leftarrow k+1$
    }
}
\Return $\bot$
\end{algorithm}
\begin{table}[hb]
\centering
\label{tab:tool_complexity}
\begin{tabular}{lcc}
\toprule
\textbf{Tool} & \textbf{Time Complexity} & \textbf{Description} \\
\midrule
\texttt{Filter} & $O(|V| + |E|)$ & Metadata filter \\
\texttt{Search} & $O(|V| \cdot d_{\text{embed}} + k \log k)$ & Vector retrieval + sorting \\
\texttt{GraphTraversal} & $O(\bar{d} \cdot \text{depth})$ & BFS ($\bar{d}$: avg degree) \\
\texttt{Temporal} & $O(1)$ & Indexed temporal query \\
\texttt{AudioAccess} & $O(1)$ & Get audio clip \\
\bottomrule
\end{tabular}
\caption{Time Complexity of Graph Operations}
\end{table}

\twocolumn

\subsection{Complexity Analysis}
We provide a comprehensive theoretical analysis of the time and space complexity of the proposed GRGA algorithm. We examine each component's computational cost and derive the overall complexity bounds.

\subsubsection{Time Complexity Analysis}

The total time complexity of GRGA is determined by the iterative POMDP loop and its constituent operations:

\begin{equation}
\label{eq:time_total}
T_{\text{total}} = T_{\text{dec}} + \sum_{k=0}^{K_{\max}} \left( T_{\text{plan}}^{(k)} + T_{\text{exec}}^{(k)} + T_{\text{syn}}^{(k)} + T_{\text{ref}}^{(k)} \right)
\end{equation}

where $K_{\max}$ is the maximum number of iterations.

\paragraph{Query Decomposition.}
The decomposition function $f_{\text{dec}}: \mathcal{Q} \to \mathcal{C}$ involves a single LLM inference:
\begin{equation}
T_{\text{dec}} = O(|Q| \cdot d_{\text{model}})
\end{equation}
where $|Q|$ denotes the query length in tokens and $d_{\text{model}}$ is the model dimension. This is a one-time operation outside the main loop.

\paragraph{Execution Planning.}
At iteration $k$, the policy network $\pi_\theta$ generates a plan conditioned on the belief state $b_k$:
\begin{equation}
T_{\text{plan}}^{(k)} = O(|b_k| \cdot d_{\text{model}} + L_{\text{plan}} \cdot |\mathcal{A}|)
\end{equation}
where:
\begin{itemize}
    \item $|b_k| = O(|Q| + |\mathcal{C}| + k \cdot |o_{\text{avg}}|)$ grows linearly with iteration count
    \item $L_{\text{plan}}$ is the average plan length
    \item $|\mathcal{A}|$ is the action space size
\end{itemize}

\paragraph{Tool Execution.}
The execution engine applies $L_{\text{plan}}$ operations on the meeting graph $\mathcal{G} = (V, E)$. The complexity depends on the tool type:

The worst-case execution time is:
\begin{equation}
T_{\text{exec}}^{(k)} = O(L_{\text{plan}} \cdot \max\{|V| \cdot d_{\text{embed}}, |V| \log |V|\})
\end{equation}

\paragraph{Answer Synthesis.}
The synthesizer $f_{\text{syn}}$ processes accumulated evidence:
\begin{equation}
T_{\text{syn}}^{(k)} = O\left(\left|\bigcup_{i=0}^{k} o_i\right| \cdot d_{\text{model}} + L_{\text{answer}} \cdot d_{\text{model}}\right)
\end{equation}
where $L_{\text{answer}}$ is the generated answer length. Note that $\left|\bigcup_{i=0}^{k} o_i\right| = O(k \cdot |o_{\text{avg}}|)$ grows with iterations.

\paragraph{Reflection.}
The reflector $f_{\text{ref}}$ evaluates logical entailment:
\begin{equation}
T_{\text{ref}}^{(k)} = O\left((|Q| + |\hat{A}_k| + |o_k|) \cdot d_{\text{model}}\right)
\end{equation}

\paragraph{Aggregate Complexity}

\begin{equation}
\label{eq:time_complexity_final}
\begin{split}
T_{\text{total}} = O\bigg( K_{\max} \cdot \big[ &(|Q| + k \cdot |o_{\text{avg}}|) \cdot d_{\text{model}} \\
&+ L_{\text{plan}} \cdot |V| \cdot d_{\text{embed}} \big] \bigg)
\end{split}
\end{equation}

Under typical conditions where $|V| \gg |Q|$ and $d_{\text{embed}} \approx d_{\text{model}}$, this simplifies to:
\begin{equation}
T_{\text{total}} = O(K_{\max} \cdot |V| \cdot d_{\text{model}})
\end{equation}

In practice, $K_{\max} \in [1, 5]$ and early stopping (when $s_{\text{ver}} \geq \tau$) significantly reduces the average iteration count $\bar{K} < K_{\max}$.

\subsection{Space Complexity Analysis}

The space requirements consist of three main components:
\begin{equation}
S_{\text{total}} = S_{\text{graph}}
\end{equation}

\paragraph{Graph Storage.}
The meeting graph requires storage for structure and embeddings:
\begin{equation}
S_{\text{graph}} = O(|V| + |E| + |V| \cdot d_{\text{embed}})
\end{equation}

For a typical 30 minutes meeting:
\begin{itemize}
    \item $|V| \approx 900$ nodes
    \item $d_{\text{embed}} = 1024$  (BGE-M3)
    \item Storage: $\sim$2MB (structure) + $\sim$3.5MB (embeddings)
\end{itemize}

\section{Tools Details}
The table~\ref{tab:tool_definitions} shows the definitions of the atomic tools in our Action Space (A).
\begin{table*}[]
    \centering
    \resizebox{\textwidth}{!}{
        \begin{tabular}{lp{5.5cm}p{7.5cm}}
            \toprule
            \textbf{Category} & \textbf{Tool Signature} & \textbf{Description \& Purpose} \\
            \midrule
            \multirow{3}{*}{\textbf{Retrieval}} & \texttt{keyword\_search(query)} & BM25-based search for precise entity/term lookup. \\
             & \texttt{semantic\_search(query)} & Dense vector retrieval for abstract semantic concepts. \\
             & \texttt{hybrid\_search(query, }$ \alpha=0.6 $\texttt{)} & Weighted combination of keyword and semantic scores to optimize recall. \\
            \midrule
            \multirow{2}{*}{\textbf{Filtering}} & \texttt{filte\_time\_range(}$ t_{start}, t_{end} $\texttt{)} & Retrieves utterances strictly within a specified time window. \\
             & \texttt{filte\_speaker(nodes, spk\_id)} & Filters a set of candidate nodes by speaker identity. \\
            \midrule
            \textbf{Traversal} & \texttt{traverse\_relations(nodes, depth=k)} & Walks along graph edges (e.g., \textit{Next}, \textit{Reply-To}) to trace dialogue threads for multi-hop reasoning. \\
            \midrule
            \textbf{Audio} & \texttt{audio\_segment(}$ t_{start}, t_{end} $\texttt{)} & Retrieves raw waveform data to ground text in acoustic signals (e.g., for emotion detection). \\
            \bottomrule
        \end{tabular}
    }
    \caption{The definitions of the atomic tools in our Action Space ($ \mathcal{A} $). The Query Planner invokes these tools to interact with the meeting graph and retrieve evidence.}
    \label{tab:tool_definitions}
\end{table*}

\clearpage
\onecolumn
\section{QA Examples}
\label{qa_example}
\paragraph{Header.}
The box title \texttt{Example~\# (QA\_TYPE, Key=\#)} indicates the example index, its question type (e.g., \categorylabel{cfactual}{Factual}), and a unique identifier \texttt{Key} used to retrieve the corresponding record in the released JSON files.

\paragraph{Evidence section (lower part).}
Below the dashed line, we display the \textbf{verbatim evidence excerpt(s)} for the referenced utterance IDs.
Each excerpt is formatted as:
\texttt{Evidence [utt\_id]: [t\_start -- t\_end] SPEAKER\_\#: transcript}.
Here, \texttt{[t\_start -- t\_end]} denotes the absolute timestamps (in seconds) within the meeting, and \texttt{SPEAKER\_\#} is the diarization-based speaker label.
This layout makes the supervision explicitly \emph{grounded}: readers can directly verify that the answer is entailed by the cited utterance(s), and systems can be evaluated on both answer correctness and evidence retrieval.

\vspace{-0.25em}

\begin{qabox}{Example 1 (\categorylabel{cfactual}{Factual}, Key=1)}
\qafield{Question}{What specific technical difficulty is Speaker 4 facing regarding XML files?}
\qafield{Answer}{Speaker 4 has written code to read and remove the XML, but is struggling to store the data (e.g., into a vector) or display it on the screen using Java.}
\qafield{Evidence (utterance IDs)}{[513]}
\qafield{Rationale}{Speaker 4 explicitly states their current coding roadblock regarding Java vectors and display.}
\tcblower
\qafield{Evidence [513]}{[1083.43 - 1100.41] SPEAKER\_4: i've been trying to g write something to read the x. m. l. and get rid of it and i can get rid of it but i'm having trouble putting it anywhere else so it will come up on the screen for the moment i haven't managed to put it into a vector or whatever in java to play with it.}
\end{qabox}
\vspace{0.6em}

\begin{qabox}{Example 2 (\categorylabel{cinferential}{Inferential}, Key=26)}
\qafield{Question}{How did the group decide to resolve the issue of having too many pop-up windows?}
\qafield{Answer}{The group discussed using tabs (similar to Mozilla) or toggle buttons within a single window to switch between views like the full transcription and the summary, rather than opening separate windows.}
\qafield{Evidence (utterance IDs)}{[265, 266, 269, 273, 304, 314]}
\qafield{Rationale}{Multiple speakers contribute to the idea of using tabs/toggles to manage content within one window frame to avoid clutter.}
\tcblower
\qafield{Evidence [265]}{
[657.56 - 663.43] SPEAKER\_2: so maybe you can just like choose the s same window for transcription and summary.
}
\qafield{Evidence [266] }{
[661.99 - 664.73] SPEAKER\_1: hmm so like have a tab there.
}
\qafield{Evidence [269] }{
[665.32 - 669.42] SPEAKER\_2: yeah yeah tabs are nice.
}
\qafield{Evidence [273] }{
[669.91 - 671.05] SPEAKER\_4: mozilla style.
}
\qafield{Evidence [304] }{
[697.23 - 703.29] SPEAKER\_2: yeah uh change the contents of the same window like from transcription to summary.
}
\qafield{Evidence [314] }{
[719.03 - 725.56] SPEAKER\_2: no no it could be like transcription summary like two buttons and you just press on which ever you want.
}

\end{qabox}
\vspace{0.6em}

\begin{qabox}{Example 3 ( \categorylabel{csummarization}{Summarization}, Key=53)}
\qafield{Question}{What are the key interface design decisions made during this meeting?}
\qafield{Answer}{The team decided to: (1) use buttons instead of right-click menus for speaker characterisation, (2) use tabs or buttons to toggle between transcription and summary in the same window rather than separate windows, and (3) implement right-click menus for topics with options to view all meetings containing that topic.}
\qafield{Evidence (utterance IDs)}{[76, 84, 88, 158, 167, 266, 269, 314]}
\qafield{Rationale}{These utterances capture the major UI/UX consensus points reached through discussion.}

\tcblower

\qafield{Evidence [76]}{[207.74 - 220.21] SPEAKER\_2: i guess a button button makes a bit more sense 'cause otherwise you don't really know that oh what if i right click now what happens then it's like more if it's visual.}
\qafield{Evidence [84]}{[225.08 - 232.04] SPEAKER\_4: it's more idiot proof isn't it it's got a button.}
\qafield{Evidence [88]}{[229.16 - 231.24] SPEAKER\_3: that's true yeah it's more intuitive really isn't it.}
\qafield{Evidence [158]}{[399.55 - 402.78] SPEAKER\_3: so we could do that in a similar way do it right click as well.}
\qafield{Evidence [167]}{[411.28 - 424.13] SPEAKER\_3: so we have basically two options of of browsing the meetings is by either um searching and opening individual observations and when then we have the interlinking by right click basically.}
\qafield{Evidence [266]}{[661.99 - 664.73] SPEAKER\_1: hmm so like have a tab there.}
\qafield{Evidence [269]}{[665.32 - 669.42] SPEAKER\_2: yeah yeah tabs are nice.}
\qafield{Evidence [314]}{[719.03 - 725.56] SPEAKER\_2: no no it could be like transcription summary like two buttons and you just press on which ever you want.}
\end{qabox}

\vspace{0.6em}

\begin{qabox}{Example 4 (\categorylabel{ctemporal}{Temporal}, Key=60)}
\qafield{Question}{At what point does the team shift from discussing GUI to discussing the interim prototype?}
\qafield{Answer}{Around 810--824 seconds (approximately 13--14 minutes into the meeting).}
\qafield{Evidence (utterance IDs)}{[361, 363, 365]}
\qafield{Rationale}{The topic pivot occurs when Speaker 3 asks what prototype they should aim for.}

\tcblower

\qafield{Evidence [361]}{[816.58 - 822.38] SPEAKER\_3: and finally the prototype he spoke about what kind of prototype could we produce.}
\qafield{Evidence [363]}{[824.27 - 828.76] SPEAKER\_3: because i'm i'm just you know i go into the lab and i say right what am i gonna change today.}
\qafield{Evidence [365]}{[828.76 - 836.48] SPEAKER\_3: you know and it kind of just it just develops i'm not aiming for anything do we wanna aim for something.}

\end{qabox}
\vspace{0.6em}

\begin{qabox}{Example 5 (\categorylabel{cacoustic}{Acoustic}, Key=\textit{80})}
\qafield{Question}{Why did Speaker 3 express sadness when saying ``well i hope so'' at approximately 24 seconds?}
\qafield{Answer}{Speaker 3 sounded sad when expressing hope that others had done some work, likely in response to Speaker 4's question ``has anybody done anything'' and Speaker 1's admission ``not a lot no'', suggesting disappointment about the team's progress.}
\qafield{Evidence (utterance IDs)}{[13, 11, 12]}
\qafield{Rationale}{This combines the sad emotion label from node 13 with the contextual text from surrounding nodes to explain the cause.}
\tcblower

\qafield{Evidence [11]}{[ 19.07 -  23.69] SPEAKER\_4: has anybody done anything.}
\qafield{Evidence [12]}{[ 21.47 -  23.29] SPEAKER\_1: not a lot no.}
\qafield{Evidence [13]}{[ 24.17 -  25.22] SPEAKER\_3: well i hope so.}
\end{qabox}

\section{Manual Evaluation Tool}

\begin{figure}[h!]
    \centering
    \includegraphics[width=1\linewidth]{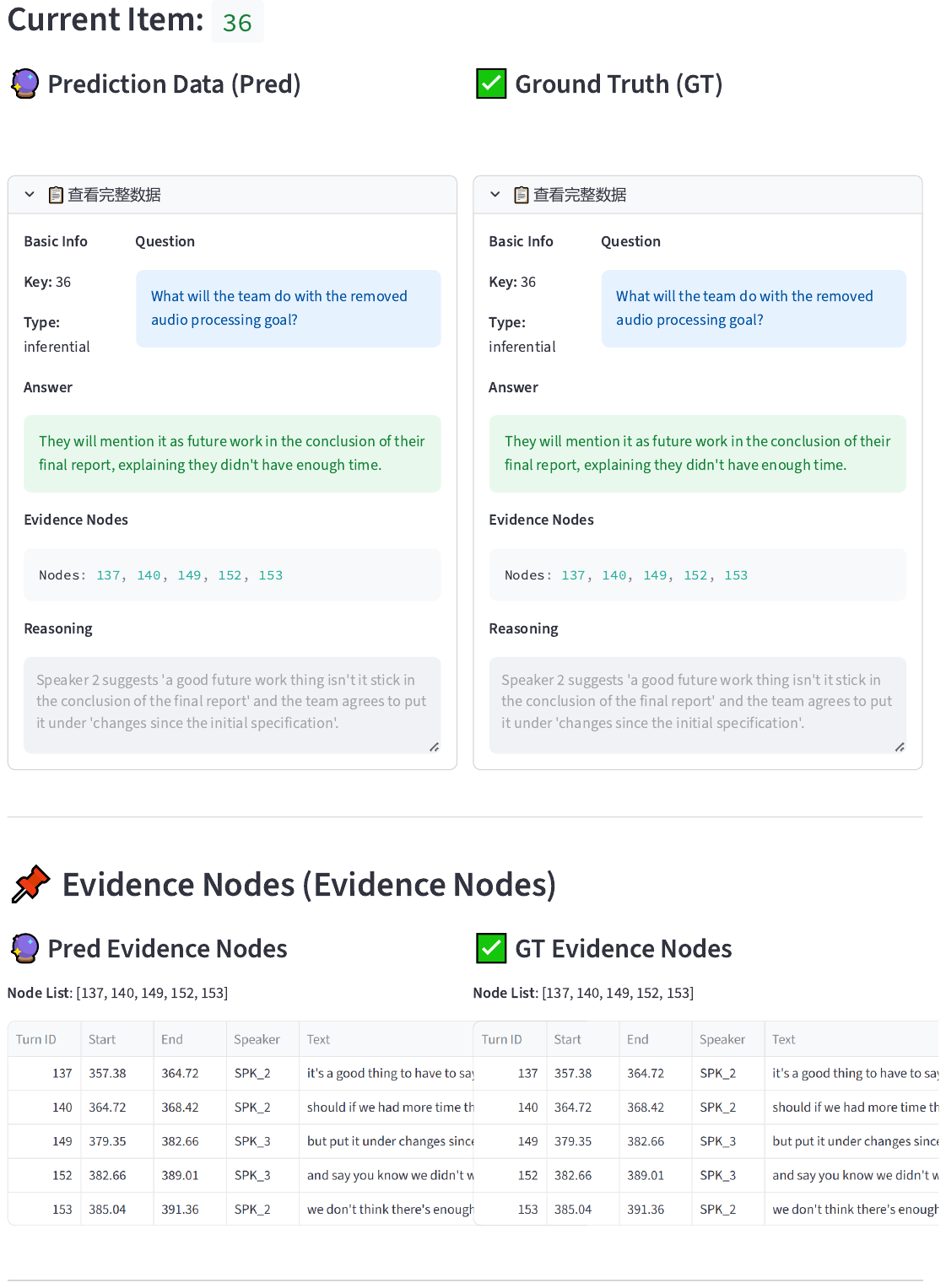}
    \caption{Data Display}
    \label{fig:Manual1}
\end{figure}

\begin{figure}[h!]
    \centering
    \includegraphics[width=1\linewidth]{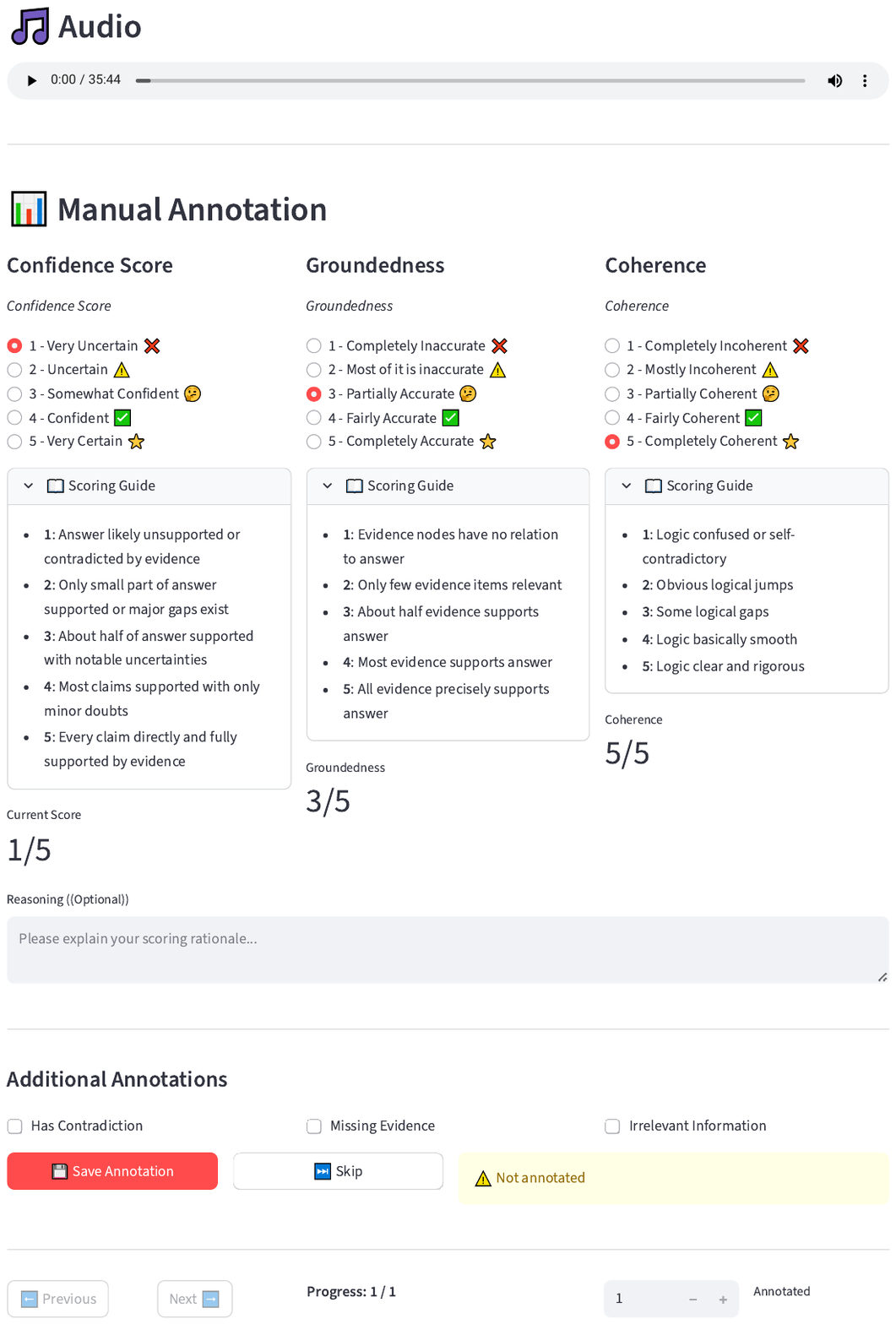}
    \caption{Manual Evaluation}
    \label{fig:Manual2}
\end{figure}

\section{Data Construction Prompt}
\begin{promptbox}[Prompt for Acoustic-Aware QA Generation]
\textbf{\# Role}\\
You are an expert AI assistant specializing in multimodal speech dataset construction. 
Your task is to generate high-quality \textbf{Acoustic-Aware QA pairs} based on the provided audio and speech transcription and metadata.

\vspace{0.5em}
\textbf{\# Input Data format}\\
An audio clip.\\
A list of Nodes. Each Node contains:\\
`turn\_id`: [`timestamp start`-`timestamp end`] Speaker `speaker`: `text` (Emotion: `emotion`) (Volume: `Volume`)

\vspace{0.5em}
\textbf{\# Task Definition: Acoustic-Aware QA}\\
You must generate questions that \textbf{cannot} be answered by reading the text alone. The answer must require checking the \textbf{Acoustic Features} (specifically the `emotion` and `timestamp` fields).

\vspace{0.5em}
\textbf{\#\# Critical Requirement 1: Focus on High Arousal Emotions}\\
You must prioritize questions regarding strong emotions such as \textbf{Anger, Happiness, Excitement, or Sadness}.

\textbf{\#\#\# How to handle "Neutral" data:}\\
If the input data contains only "Neutral" labels:
\begin{itemize}[leftmargin=1.5em, nosep]
    \item \textbf{Do:} Ask \textbf{Verification Questions} checking for the presence of strong emotions.
    \begin{itemize}[leftmargin=1.5em]
        \item \textit{Good Example:} "Did Speaker 3 sound \textbf{angry} when asking about the annual meeting?"
        \item \textit{Good Answer:} "No. According to the audio data, Speaker 3 maintained a \textbf{neutral} tone."
    \end{itemize}
    \item \textbf{Don't:} Ask passive descriptive questions like \textit{"What was the emotion?"}.
\end{itemize}

\vspace{0.5em}
\textbf{\#\# Critical Requirement 2: NO Confidence Scores}
\begin{itemize}[leftmargin=1.5em, nosep]
    \item \textbf{Strictly Forbidden:} Do \textbf{NOT} mention the numeric confidence scores (e.g., `0.99`, `1.0`) in the Question or the Answer.
    \item Simply state the emotion label as a fact (e.g., "The speaker was angry").
\end{itemize}

\vspace{0.5em}
\textbf{\#\# Allowed Question Patterns}
\begin{enumerate}[leftmargin=2em, nosep]
    \item \textbf{Emotion-Content Attribution (Why is he angry?):}\\
    "Why did spk3 sound \textbf{angry} at the 15th minute?" (Combine Emotion label + Text analysis).
    \item \textbf{Specific Time Identification:}\\
    "Who sounded the most \textbf{excited} around 10 seconds?"
    \item \textbf{Emotion Verification (For Neutral Data):}\\
    "When Speaker 1 said [Text], did they express \textbf{happiness}?"\\
    "Did the speaker sound \textbf{furious} or \textbf{annoyed} during the discussion?"
\end{enumerate}

\vspace{0.5em}
\textbf{\#\# NOT Allowed Question Patterns}
\begin{itemize}[leftmargin=1.5em, nosep]
    \item \textbf{Wrong Questions:}\\
    Q: Why did Speaker 3 sound sad when saying 'i think the search works as well' around 45 seconds?\\
    A: Speaker 3 did not sound sad at that moment.
\end{itemize}

\vspace{0.5em}
\textbf{\# Output Format (Strict JSON)}\\
Output a valid JSON List. Ensure `evidence\_nodes` is a list of integers. Using ' instead of " for citation

\begin{verbatim}
```json
[
  {{
    "question": "The specific 'question' string.",
    "answer": "The brief 'answer' derived from the text.",
    "evidence_nodes": [10, 12], // Must be a list of turn_ids
    "reasoning": "Brief explanation of the logic."
  }}
]
```
\end{verbatim}
\vspace{0.5em}
\textbf{\# Input:}
\begin{verbatim}
{input_data}
\end{verbatim}

\textbf{\# Output:}
\end{promptbox}

\begin{promptbox}[Prompt for Graph-based QA Generation]
\textbf{\# Role}\\
You are an expert data annotator specialized in \textbf{Graph-based Audio Understanding} and \textbf{Conversational AI}. 
Your goal is to construct a massive, high-quality Graph RAG dataset from raw ASR transcripts.

\vspace{0.5em}
\textbf{\# Input Data Structure}\\
A list of Nodes. Attributes: \\
`turn\_id`: [`timestamp start`-`timestamp end`] Speaker `speaker`: `text`

\vspace{0.5em}
\textbf{\# Mission}\\
Your mission is to \textbf{exhaustively} mine the provided transcript for every possible piece of information and convert it into a Question-Answer (QA) pair. \textbf{Do not stop at just one question per category.} Aim to generate as many valid, distinct QA pairs as the data supports (e.g., 10-20 pairs for a medium-length segment).

\vspace{0.5em}
\textbf{\# Critical Guidelines (The "Don't"s)}
\begin{enumerate}[leftmargin=2em, nosep]
    \item \textbf{Filter Noise:} Do NOT generate Factual questions about "YEAH", "OKAY", "MM-HMM" (backchannels).
    \begin{itemize}[leftmargin=1.5em]
        \item \textit{Bad:} "Who said 'Yeah'?"
        \item \textit{Good (Inferential):} Use "Yeah" nodes only as evidence for "Agreement" or "Consensus".
    \end{itemize}
    \item \textbf{Avoid Hallucination:} Every answer must be strictly supported by the `evidence\_nodes`.
    \item \textbf{No Generic Questions:} Avoid vague questions like "What happened?". Be specific: "What software bug did Speaker A mention?"
\end{enumerate}

\vspace{0.5em}
\textbf{\# Question Generation Strategy (How to generate MANY questions)}\\
To maximize the number of QAs, apply these specific strategies:

\vspace{0.5em}
\textbf{\#\# Strategy A: Entity-Centric Mining (Factual)}\\
Scan for \textbf{every} entity in the text. Generate a question for each.
\begin{itemize}[leftmargin=1.5em, nosep]
    \item \textbf{Entities:} Project names, specific numbers, technical terms, people's names, locations, file paths.
    \item \textit{Trigger:} "I see a phone number '04555'." -> \textit{QA:} "What is the phone number mentioned?"
\end{itemize}

\vspace{0.5em}
\textbf{\#\# Strategy B: Logic \& Linkage (Inferential)}\\
Look for \textbf{connections} between distant or adjacent nodes.
\begin{itemize}[leftmargin=1.5em, nosep]
    \item \textbf{Cause \& Effect:} Node X proposes something -> Node Y rejects it. -> \textit{QA:} "Why was the proposal in Node X rejected?"
    \item \textbf{Clarification:} Node A asks a question -> Node B answers it. -> \textit{QA:} "How did Speaker B respond to A's inquiry about [Topic]?"
    \item \textbf{Sentiment:} Look for emotional words or emphatic language. -> \textit{QA:} "Which speaker expressed frustration about the file system?"
\end{itemize}

\vspace{0.5em}
\textbf{\#\# Strategy C: Time Anchors (Temporal)}
\begin{itemize}[leftmargin=1.5em, nosep]
    \item \textbf{Absolute:} "What topic was introduced exactly at the 10-minute mark?"
    \item \textbf{Relative:} "What was discussed \textit{immediately before} the discussion about the budget?"
    \item \textbf{Duration:} "How long did the debate about 'UI Design' last?" (Calculate from start/end timestamps of the cluster).
\end{itemize}

\vspace{0.5em}
\textbf{\#\# Strategy D: Scene Understanding (Summarization)}
\begin{itemize}[leftmargin=1.5em, nosep]
    \item \textbf{Topic Segmentation:} Identify where the topic shifts. -> \textit{QA:} "Summarize the main points discussed regarding [Topic Name]."
    \item \textbf{Speaker Role:} "Based on the segment, what seems to be the role of Speaker 0?" (e.g., Manager, Technical Lead).
\end{itemize}

\vspace{0.5em}
\textbf{\# Output Format (Strict JSON)}\\
Output a valid JSON List. Ensure `evidence\_nodes` is a list of integers.

\begin{verbatim}
```json
[
  {{
    "type": "factual | inferential | temporal | summarization",
    "question": "The specific question string.",
    "answer": "The precise answer derived from the text.",
    "evidence_nodes": [10, 12], // Must be a list of turn_ids
    "reasoning": "Brief explanation of the logic."
  }}
]
```
\end{verbatim}

\vspace{0.5em}
\textbf{\# Input:}\
\begin{verbatim}
{input_data}
\end{verbatim}

\vspace{0.5em}
\textbf{\# Output:}
\end{promptbox}
\end{document}